\documentclass[notitlepage,pra,twocolumn,showpacs,amsmath,amssymb,showkeys]{revtex4-1}[11]

\usepackage{graphicx}
\usepackage{dcolumn}
\usepackage{bm}
\usepackage{wrapfig}
\usepackage{graphicx}
\usepackage{xcolor}
\usepackage[latin1]{inputenc}
\usepackage{graphicx}
\usepackage{graphics}
\usepackage{amsmath}
\usepackage{amssymb}
\usepackage{amsfonts}
\usepackage{mathrsfs}
\usepackage{latexsym}
\usepackage{hyperref} 
	\hypersetup{
    		colorlinks = true,
    		linkbordercolor = {white},
			}

\makeatletter

\begin{document}
\author{L. Bellino}
\affiliation{Dipartimento di Meccanica, Matematica e Management, Politecnico di Bari, Via E. Orabona 4, I-70125 Bari, Italy}

\author{G. Florio}
\affiliation{Dipartimento di Meccanica, Matematica e Management, Politecnico di Bari, Via E. Orabona 4, I-70125 Bari, Italy}
\affiliation{INFN, Sezione di Bari, I-70126 Bari, Italy}

\author{G. Puglisi}
\affiliation{Dipartimento di Scienze dell'Ingegneria Civile e dell'Architettura, Via Re David 200, 700126, Politecnico di Bari, Italy}

\date{\today}

\title{On the influence of device handle in single-molecule experiments}

\begin{abstract}
\noindent We deduce a fully analytical model to predict the artifacts of the measuring device handles in Single Molecule Force Spectroscopy experiments. As we show,  neglecting the effects of the handle stiffness  can lead to crucial overestimation or underestimation of the stability properties and transition thresholds of macromolecules.  
\end{abstract}

%
\maketitle
\section{Introduction}

\begin{figure*}
\centering
  \includegraphics[width=0.95\textwidth]{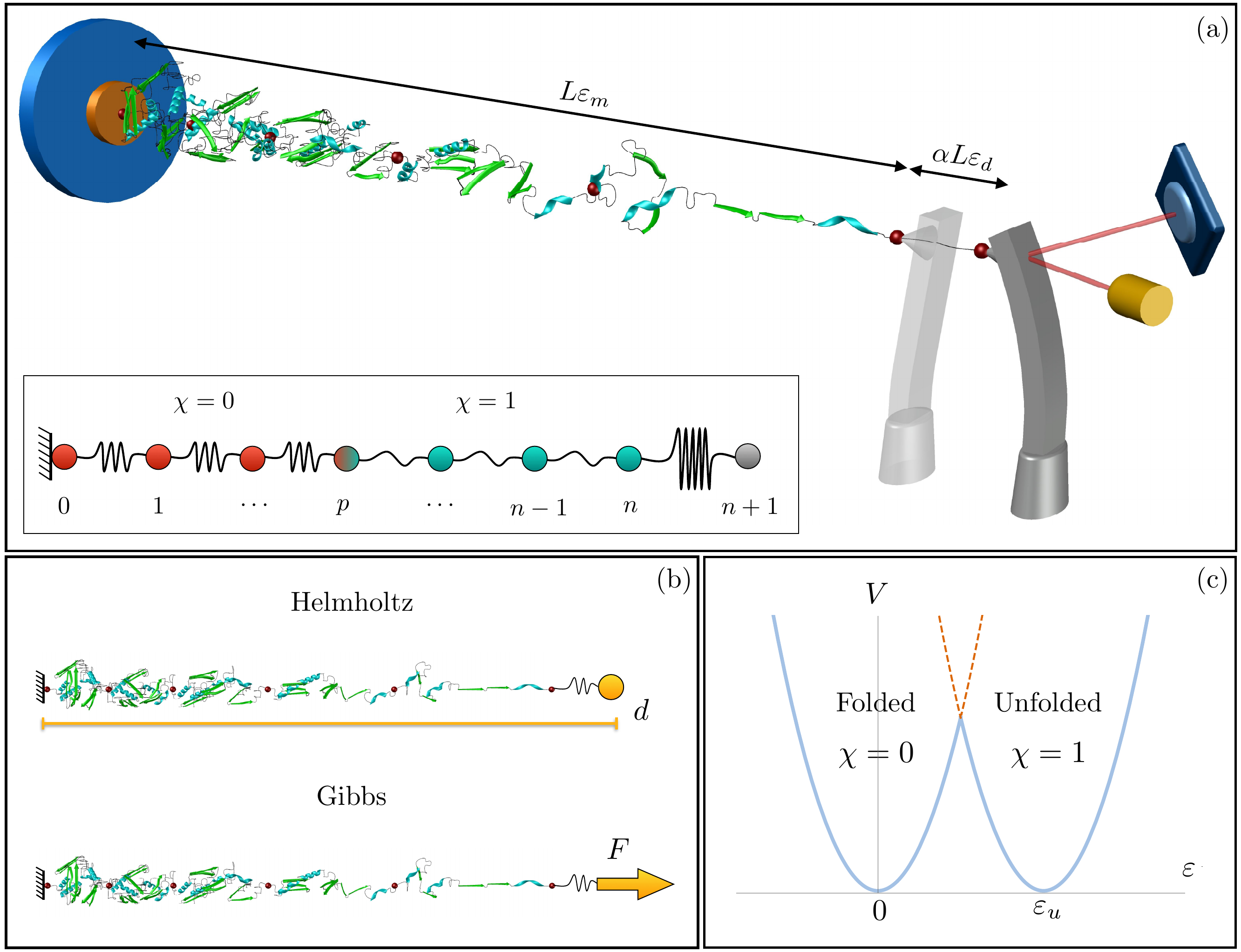}
  \caption{Model. (a) Scheme of a SMFS  unfolding experiment of a molecule of domains undergoing folded$\rightarrow$unfolded transition and of the considered mechanical system (inset). (b) Boundary conditions: applied displacement (Helmholtz) or applied force (Gibbs). (c) Two-wells elastic energy of the elements.}
  \label{fig:model}
\end{figure*}

The mechanical response of molecules represents a crucial topic in Biomechanics, Biology, Medicine and Material Engineering~\cite{goriely:2017}.  As explicit examples we may refer to the morphogenesis of the neuronal network~\cite{recho:2016}, to the influence of contour instabilities in tumor growth~\cite{benamar:2011}, to the misfolding of the $\alpha$-synuclein protein in Parkinson's disease~\cite{krasnoslobodtsev:2013}, to the gene transcription in DNA~\cite{gonzalez:2017,woodside:2008} and to folding/unfolding processes in RNA~\cite{li:2008}. 

The breakthrough that opened up the possibility of investigating the force response at the molecular level can be addressed to Single Molecule Force Spectroscopy (SMFS) techniques. An outstanding innovation in the last decades delivered high-precision instruments such as Atomic Force Microscopes (AFM), optical tweezers, magnetic tweezers, and micro-needles~\cite{bustamante:2000} letting the possibility of unreached precision micro and nano scale mechanical experiments. Being characterized by very different operational ranges of spatial resolution, stiffness, displacement and probe size, the choice of the instrument depends on the specific application~\cite{neuman:2008}. For instance, AFM experiments are adopted for macromolecule pulling and interaction tests~\cite{dutta:2016,hughes:2016} whereas optical tweezers are better suited for DNA synthesis analysis~\cite{woodside:2006}. 

The deduction of detailed properties of the complex energy landscape of macromolecules is probably one the most interesting opportunity allowed by SMFS experiments. This is due to the possibility of considering different loading directions and rates, enabling the analysis of the relative stability of the multiple metastable configurations~\cite{szabo:2005}. Nevertheless, to attain effective results, several fundamental questions still need to be analyzed, going from a detailed definition of the force direction~\cite{walder:2017} to the analysis of loading rate effects~\cite{benichou:2016,cossio:2015}.

However, the most important issue, on which this paper is focussed, concerns the strong influence of the handling device stiffness on the experimental response~\cite{maitra:2010}, an aspect often underestimated or even neglected. Indeed, as it will be clear in the following, the finite value of the cantilever stiffness of an AFM can induce a wrong estimation of both the dissipated energy and the unfolding force thresholds, leading to important discrepancies between the theoretical {\it vs} experimental forces~\cite{biswas:2018} and transition rates~\cite{dudko:2008,li:2014}. Indeed, since the adopted SMFS devices are characterized by a device stiffness differing of several order of magnitude~\cite{bustamante:2000}, the analysis of this effect cannot be neglected. 

Our aim is the deduction of an analytical description, in the framework of equilibrium Statistical Mechanics~\cite{weiner:1983}, of the experimental device influence in the whole range of its stiffnesses on a macromolecule undergoing a two-states  transition. While our approach and analytical results are general, as a paradigmatic example we refer to the fundamental case of AFM induced unfolding experiments in bio-macromolecules such as titin~\cite{rief:1997,givli:2011}. In this case the molecules are characterized by domains undergoing a conformational (folded $\rightarrow$ unfolded) transition~\cite{rief:2002}. We remark that our model is restricted to the rate of loading such that the behavior is rate-independent \cite{chung:2013}. Kramer's type approaches have been adopted to deduce that in the rate-dependent regime the unfolding force logarithmically grows with the rate of loading\cite{suzuki:2013}.

Specifically, following~\cite{detom:2013,manca:2013,benichou:2013,makarov:2009,bustamante:2003}, in this paper we consider an energy-based approach where we describe the unfolding macromolecule as a chain of bistable elements~\cite{puglisi:2000} with a two-wells elastic energy (called {\it folded and unfolded states}) and,  to get explicit analytical results, we assume that the element behavior inside each well is Gaussian. This approximation is acceptable as far as the extension of the molecule is much shorter than the contour length both in the folded and unfolded configuration. Of course more general models can be considered in accordance with the molecule under investigation, but this would allow only for numerical results. As important examples we recall that a Freely Jointed Chain (FJC) energy was chosen in~\cite{su:2009,benedito:2018} where different phases are characterized by different Kuhn lengths, whereas a Worm Like Chain (WLC) description of the unfolded configuration with a stiffer energy of the folded state has been considered in~\cite{detom:2013,staple:2008}.  

In this paper we study the different behavior of the introduced chain as the boundary conditions, induced by the loading device, vary. In passing it may be interesting to observe that the following analysis of the interaction with the handling system can elucidate important aspects regarding the elastic interaction between different molecules.

Two \textit{ideal} limit regimes can be considered (see Fig.~\ref{fig:model}b). In the \textit{ideal hard device}, the chain elongation is assigned and the force is a fluctuating conjugated variable. In this case, as the extension is increased, a sequence of localized transitions occur with a typical sawtooth force-elongation diagram. In the opposite hypothesis, here called \textit{ideal soft device}, isotensional experiments are considered. A force is applied at the last element of the chain and the overall elongation represents the conjugate unknown variable. In this case the transition path is monotonic with a more cooperative transition behavior. As remarked above, this peculiar feature characterizes the behavior of stretching experiments of other multistable systems:  deformations localization and sawtooth transition paths in metal plasticity (Portevin-LeChatelier effect)~\cite{froli:2000}, in shape memory materials~\cite{puglisi:2000} and in muscle contraction mechanics~\cite{hudson:2019}. 
 
For systems characterized by convex energies the analysis of the hard and soft device cases represents a simple problem in the field of Statistical Mechanics~\cite{weiner:1977}. Conversely, the problem of a multistable material is a more subtle and only partly solved subject. In this framework, a chain of bistable elements has been analyzed in~\cite{efendiev:2010}, where the authors model the thermalization of a Fermi Pasta Ulam system. Semi-analytical results are obtained in both soft and hard devices, yet neglecting the device stiffness effect. In~\cite{mancagiordano:2014} a chain of links with three-parabolic energy wells has been considered and the authors show the equivalence --in terms of mechanical response-- of the Gibbs (soft device) and Helmholtz (hard device) ensembles in the thermodynamic limit. 

On the other hand, just because of the influence of the device stiffness, real SMFS experiments  live \textit{in between} the two ideal hard and soft device limits typically considered in the literature. The analysis of this effect was  firstly elucidate in~\cite{kreuzer:2001}, where the interaction of a macromolecule with convex energy loaded by a device with variable stiffness has been studied. Instead,  the case of non-convex energy was recently analyzed in~\cite{fp:2019}, where the influence of the device stiffness in the case of assigned displacement acting on the system has been deduced. The analytical results are in very good agreement with the observations in~\cite{zhanget:2008} where the authors experimentally shows the strong influence of the pulling device on the P-Selectin molecule stretching response. 

The aim of this paper is to extend the approach in~\cite{fp:2019} and describe, in a fully analytical and self-consistent model, all the possible experimental boundary conditions. In more detail, a typical SFMS experiment can be performed in two different ways depending on the specific instruments or technology used. In particular an elastic interaction must be considered for real experiments, because the displacement ({\it hard device}) or the force ({\it soft device}) are applied to an ancilla macromolecule or to a microcantilever that is then attached to the molecule.
Usually, in AFM pulling tests one of the free ends of the macromolecule is fixed (see Fig.~\ref{fig:model}) whereas the other one is attached to an elastic handle that is subjected to a fixed displacement~\cite{rief:1997}.  On the other hand, in other cases such as magnetic and optical tweezers, the generated field is used to apply a force to the handle~\cite{kim:2009}. 

According with previous description, following~\cite{fp:2019}, where only the case of assigned displacement has been analyzed, we consider the macromolecule and the device as a single thermodynamical system and extend the results also to the case of assigned force. As we show, the Gibbs ensemble can be obtained from the Helmholtz one by a Laplace transform~\cite{weiner:1983}, also in the case when the device influence and non-convex energies are considered. Furthermore, we obtain that the ideal hard (soft) device can be deduced as limit regimes when the device stiffness is much larger (smaller) than the molecule stiffness. Finally, we show the equivalence of the two ensembles in the thermodynamic limit. This result was analytically shown in~\cite{mancagiordano:2014} for the convex Freely Jointed Chain energy whereas it was numerically deduced for the non-convex case. Here we extend this result by proving this equivalence even in the non-convex energy model,  both when the device is considered or not.

The paper is organized as follows. In Section \ref{sec:model} we present the model used throughout the manuscript. In Section \ref{sec:zero} we analyze the mechanical (zero temperature) limit. In Section \ref{sec:temp} we include the effects of temperature and consider the cases of fixed force (Gibbs ensemble) and fixed displacement (Helmholtz ensemble). In Section \ref{sec:tl} we consider the thermodynamic limit and show that the results in the two ensembles coincide. In Section \ref{sec:discussion} we provide a complete discussion about  all results  of this work and we summarize them in a close and simple analytical description.

Finally, for the  reader convenience in the Supplementary Information we report all the analytical details because we believe that this paper can also represent a compact reference for the readers interested in the application of Statistical Mechanics at systems with non convex energies under general boundary conditions. 

\section{Model}\label{sec:model}

In order to describe the (typically all or none) folded $\rightarrow$ unfolded element conformational transition, we model the macromolecule as a chain of $n$ bistable springs with reference length $l$ and total reference length $L = n l$.  Each spring has a biparabolic energy with the further assumption of identical wells with stiffness $k_m$. After introducing the `spin' variable $\chi_i$, such that  $\chi_i=0$ in the folded state and $\chi_i=1$ in the unfolded one, the total elastic energy of the macromolecule can be written as
\begin{equation}
V_m=\sum_{i=1}^{n}\frac{1}{2}\,k_m l (\varepsilon_i-\varepsilon_u\chi_i)^2,
\label{eq:1}
\end{equation}
where $\varepsilon_i$ is the strain of the $i$-th element and $\varepsilon_u$ is the unloaded strain of the second well (see Fig.~\ref{fig:model}c). 

As anticipated, the key feature of the proposed approach is that an effective analysis of the influence of the loading device on the macromolecular behavior requires to consider the macromolecule and the device as a whole thermodynamical system. Following~\cite{fp:2019}, the device influence is described by an auxiliary spring with variable stiffness $k_d$, reference length $\alpha L$, strain $\varepsilon_d$ and energy
\begin{equation}
V_d = \frac{1}{2}\alpha L \, k_d \,\varepsilon_{d}^{2}.
\label{eq:2}
\end{equation}

Moreover,  due to previous discussion, we need to introduce the {\it total elongation} (molecule plus handle) 
\begin{equation}
d=\sum_{i=1}^{n}\frac{L}{n}\,\varepsilon_i+\alpha L \varepsilon_d = L(\varepsilon_m+\alpha \, \varepsilon_d),
\label{eq:3}
\end{equation}
where
\begin{equation}
\varepsilon_m = \frac{1}{n}\sum_{i=1}^{n}\varepsilon_i
\label{eq:4}
\end{equation}
is the macromolecule's average strain. Similarly, by using eq. (\ref{eq:3}) and (\ref{eq:4}), we introduce the {\it total  averaged strain}
\begin{equation}
\varepsilon_t =\frac{d}{L(1+\alpha)} \Rightarrow (1+\alpha)\,\varepsilon_t = \varepsilon_m + \alpha \varepsilon_d.
\label{eq:5}
\end{equation}
Here and in the following we use the index $m$ to denote the macromolecule, $d$ to denote the device and $t$ to denote the total (device plus macromolecule) system quantities. Finally, we need to introduce the {\it total elastic energy} of the system 
\begin{equation}
V_t = V_m + V_d = \sum_{i=1}^{n}\frac{1}{2}\,k_m l (\varepsilon_i -\varepsilon_u\chi_i)^2 + \frac{1}{2}\alpha L \, k_d \varepsilon_{d}^{2}.
\label{eq:6}
\end{equation}
In the following we first consider the case when entropic energy terms can be neglected and then we extend the results to the general case measuring temperature effects.

\section{Mechanical Limit}\label{sec:zero}

With the aim of getting physical insight in the introduced model, we begin by considering the simple case  where thermal effects can be neglected. 

As anticipated, we consider two different boundary conditions (see Fig.\ref{fig:model}b). In one case, denoted as \textit{hard device}, we suppose that a fixed total displacement $d$ is applied  and we solve the constrained problem
\begin{equation}
\min_{\vspace{-0.1 cm}\scriptsize \begin{array}{c}{\varepsilon_{1},\dots,\varepsilon_n,\varepsilon_d}\vspace{-0.15 cm}\\ \sum_{i}\varepsilon_i/n+ \alpha \varepsilon_d =\varepsilon_t(1+\alpha)\end{array}} V_t(\varepsilon_{1},\dots,\varepsilon_n,\varepsilon_d).
\label{eq:7}
\end{equation}
In the other case, known as \textit{soft device}, a fixed constant force is applied to the free end of the chain and we search for the minima of the total potential energy:
\begin{equation}
\min_{\varepsilon_{1},\dots,\varepsilon_n,\varepsilon_d}\,\, G_t(\varepsilon_{i},\varepsilon_d) = V_t(\varepsilon_{i},\varepsilon_d) - F \left(l\sum_{i=1}^n \varepsilon_i+ \alpha L \varepsilon_d\right).
\label{eq:8}
\end{equation}

In both cases, equilibrium requires a constant force $F$ such that
\begin{equation}
\varepsilon_i = \frac{F}{k_m}+ \varepsilon_u\chi_i \quad \text{with} \quad i = 1, \dots, n \quad \text{and} \quad \varepsilon_d = \frac{F}{k_d}.
\label{eq:9}
\end{equation}
Due to the absence of non-local interactions, the equilibrium force and energy only depend on the number $p$ of unfolded elements, here assigned by the unfolded fraction 
\begin{equation}
\bar{\chi}=\sum_i \frac{\chi_{i}}{n}=\frac{p}{n}\in [0,1].
\label{eq:pn}
\end{equation}
In particular, $\bar \chi=0$ and $\bar \chi=1$ correspond to the initial fully folded state and to the fully unfolded state, respectively.  Thus, by using~\eqref{eq:4},~\eqref{eq:5} and~\eqref{eq:9} we obtain a compact expression for the equilibrium force 
\begin{equation}
F=k_m\gamma\left((1+\alpha)\,\varepsilon_t-\varepsilon_u\bar{\chi}\right)
\label{fepsilont}
\end{equation}
and for the equilibrium strain of the macromolecule
\begin{equation}
\varepsilon_m = \varepsilon_u\bar{\chi}
+ \gamma((1+\alpha)\,\varepsilon_t-\varepsilon_u\bar{\chi}).
\label{epsilontm}
\end{equation}
We introduced here the main non-dimensional parameter of the model 
\begin{equation}
\gamma=\frac{k_d}{k_d+\alpha k_m}\quad\text{with}\quad \gamma \in \,\,]0,1[ .
\label{gamma}
\end{equation}
Finally, by using \eqref{fepsilont} and \eqref{epsilontm}, we obtain the force-strain relations of the macromolecule for the equilibrium branches with different unfolded elements $p$;
\begin{equation}
F=k_m \left(\varepsilon_m - \varepsilon_u\bar{\chi}\right).
\label{Fo}
\end{equation}

Observe that by using \eqref{fepsilont} and \eqref{epsilontm}  the  two-phases equilibrium branches ($\bar{\chi}\in]0,1[$) are defined only for $|F|/k_m\le \varepsilon_u$ corresponding to a strain domain%
\begin{equation}
\frac{(1+\alpha)}{\varepsilon_u} \varepsilon_t\in 
\left \{ \begin{array}{ll} 
\displaystyle\left( -\infty , \,\,\frac{1}{\gamma}\right )  &  \mbox{ if } \quad \bar \chi=0  \vspace{0.2 cm}\\
\displaystyle\left(\bar{\chi}-\frac{1}{\gamma}, \,\, \bar{\chi}+\frac{1}{\gamma}\right) & \mbox{ if } \quad \bar \chi\in \,\, ]0,1[ \vspace{0.2 cm}\\
\displaystyle\left( 1-\frac{1}{\gamma}, \,\,+\infty \right)&  \mbox{ if } \quad \bar \chi=1
\end{array} \right..
\label{brancheseq}
\end{equation}
Moreover, due to the convexity of the wells, these solutions are locally stable in the case of both assigned force and displacement. 

\begin{figure*}
	\centering
	 \includegraphics[width=0.95\textwidth]{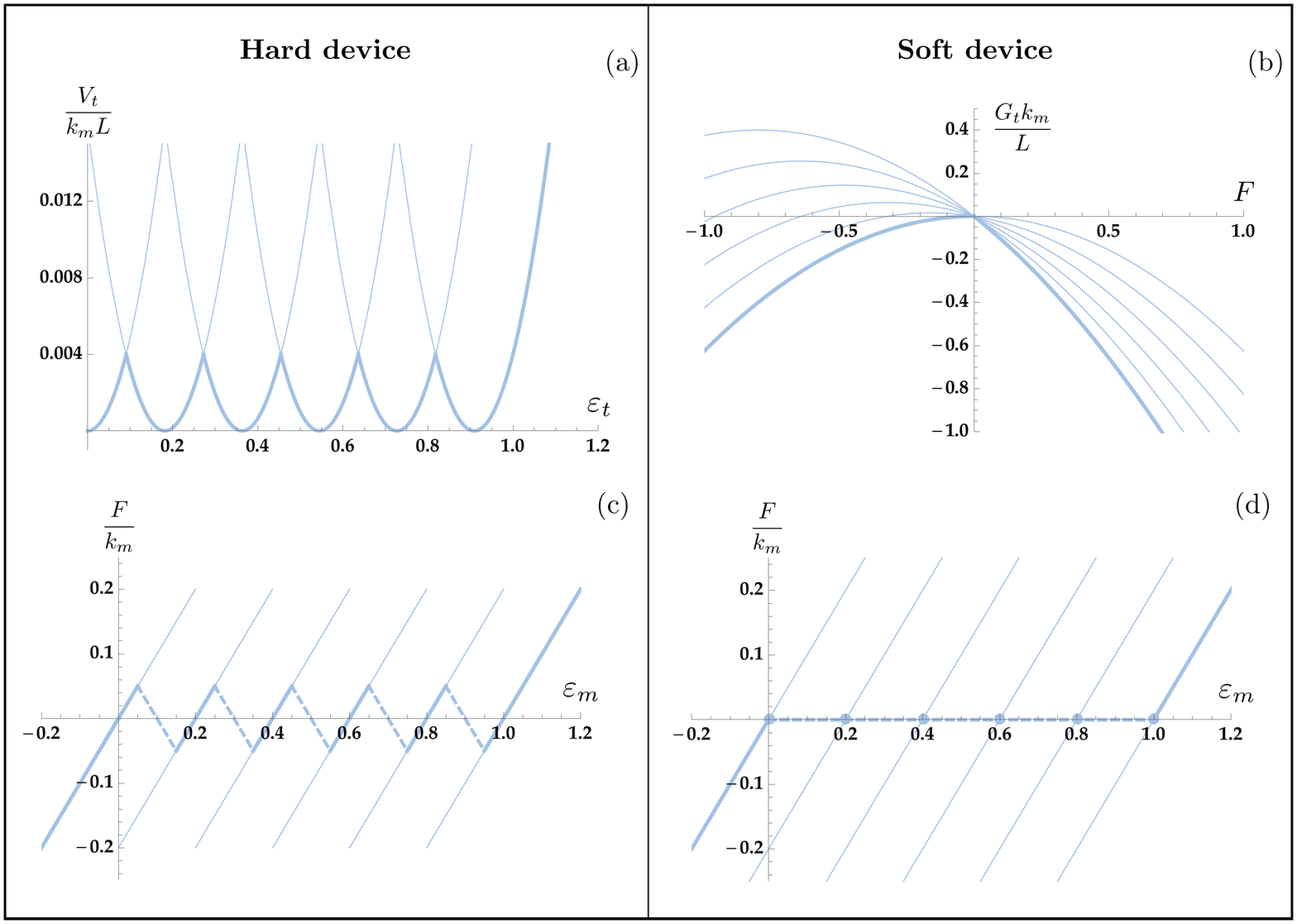}
	  \caption{Equilibrium branches (thin lines) and global energy minima (thick lines) in the mechanical limit.  (a) Total elastic energy (hard device) (a) and potential energy (b)(soft device).  Force-extension diagrams under the Maxwell convention in the hard (c) and (d) soft device: dashed lines represent strain discontinuities. Here  $\alpha=0.1$, $n=5$ and $\gamma = 0.5$, $L=1$, $k_m=1$.}
	\label{fig:energy}
\end{figure*}

To obtain the global minima of the energy we have to distinguish the two cases of hard and soft device and minimize with respect to the remaining variable $\bar \chi$. In the case of assigned displacement, for the equilibrium solutions the total elastic energy \eqref{eq:7} can be rewritten as
\begin{equation}
V_t(\varepsilon_t)=k_m L\frac{\gamma}{2}\left((1+\alpha)\varepsilon_t-\varepsilon_u\bar{\chi}\right)^2
\label{pot}
\end{equation}
with respect to the phase fraction. One can easily show that the branch  $\bar{\chi}$ corresponds to the global minimum for (see Fig.~\ref{fig:energy}a)
\begin{equation}
\frac{(1+\alpha)}{\varepsilon_u}\varepsilon_t\in 
\left \{ \begin{array}{ll} 
\displaystyle\left( -\infty , \,\,\frac{1}{2n}\right)  &  \mbox{ if } \quad \bar \chi=0  \vspace{0.2 cm}\\
\displaystyle\left(\bar{\chi}-\frac{1}{2n}, \,\, \bar{\chi}+\frac{1}{2n}\right) & \mbox{ if } \quad \bar \chi\in \,\, ]0,1[ \vspace{0.2 cm}\\
\displaystyle\left( 1-\frac{1}{2n}, \,\,+\infty \right)&  \mbox{ if } \quad \bar \chi=1
\end{array} \right..
\label{globaleq}
\end{equation}

Consequently, the force-extension behavior is assigned by \eqref{Fo} with the phase fraction depending on the total assigned strain as follows:
\begin{equation}
\bar \chi=\bar \chi(\varepsilon_t) =
\left \{ \begin{array}{ll}
\displaystyle0 &\varepsilon_t<\varepsilon_t^f \vspace{0.2 cm}\\
\displaystyle \sum_{p=0}^n\frac{p}{n}\,\, {\bf 1}_{\Omega_p} (\varepsilon_t) &\varepsilon_t^f<\varepsilon_t<\varepsilon_t^u  \vspace{0.2 cm}\\
\displaystyle1 &\varepsilon_t> \varepsilon_t^u 
\end{array} \right . 
\end{equation}
where we indicated with ${\bf 1}_{\Omega_p}$ the characteristic function of the set
\begin{equation}
\Omega_p=\left 
(\frac{\varepsilon_u}{1+\alpha}\, \left (\frac{p}{n}-\frac{1}{2n}\right ), \frac{\varepsilon_u}{1+\alpha}\, \left(\frac{p}{n}+\frac{1}{2n}\right ) \right ),
\end{equation}
 and
\begin{equation}
\varepsilon_t^f=\frac{\varepsilon_u}{1+\alpha}\, \frac{1}{2n}, \quad \quad
\varepsilon_t^u=\frac{\varepsilon_u}{1+\alpha}\left(1-\frac{1}{2n}\right ).
\end{equation}

Differently, in the case of assigned force we have to minimize the potential energy with respect to the phase fraction $\bar \chi$; in particular, for the equilibrium solutions~\eqref{eq:8} the energy can be written as
\begin{equation}
G_t(F)=-\frac{L}{2 k_m \gamma}F^2-\varepsilon_u L \bar{\chi} F.
\label{gibbsen}
\end{equation}

Thus, the global energy minimum corresponds to the fully folded state $\bar \chi=0$ for $F<0$ and to the fully unfolded state $\bar \chi=1$ for $F>0$, as shown in Fig.~\ref{fig:energy}b. Finally, the force-displacement relation is again given by \eqref{Fo} with %
\begin{equation}
\bar \chi=\bar \chi(F)={\bf 1}_{]0,\infty[}(F).
\label{sdml}
\end{equation}

The behavior of the system under the hypothesis that its configurations correspond to the global minima of the energy (Maxwell convention) are represented in Fig.~\ref{fig:energy}c with thick lines for the case of hard device. As the figure shows the transition corresponds to a sawtooth path with  the elements unfolding one at a time at a constant transition force $F=F_{un}$ that using \eqref{fepsilont} and \eqref{globaleq}  is given by 
\begin{equation}
F_{un}=\frac{k_m\,\varepsilon_u\gamma}{2n}.
\end{equation}

This behavior reflects the experimental results of the behavior of AFM unfolding experiments~\cite{rief:1997} with a periodic sawtooth path corresponding to the successive transition of the single domain. 

The case of assigned force is represented in Fig.~\ref{fig:energy}d. Observe that under this boundary conditions the transition is always cooperative, with a single value force threshold independent on the relative stiffness parameter $\gamma$.

It is important to remark that the experiments show both in the case of hard and soft device a hardening behavior with the unfolding force increasing with the unfolded fraction~\cite{rief:1997}. Interestingly, in the following we show that this hardening behavior can be associated to an entropic effect.

The main point that we can already observe in the mechanical limit is the strong dependence of the stability domains and of the unfolding force on the device stiffness, with a linear dependence of the force on both $\gamma$ and the discreteness parameter $n$. While we can already deduce that the behavior of the system in the hard device reproduces the behavior of the soft device in both cases of $\gamma \rightarrow 0$ and $n\rightarrow \infty$, we postpone this discussion to Section~\ref{sec:tl}. There, we  obtain analytically  this  new result even in the case when we do not neglect entropic energy terms.

\section{Temperature effects: Helmholtz and Gibbs statistical ensembles}\label{sec:temp}

In this section we analyze the temperature effects in the case of hard and soft devices, corresponding, respectively, to the Helmholtz and Gibbs ensembles in the framework of  Statistical Mechanics. Thus, as in the mechanical limit, we consider the system \emph{and} the measuring device as a whole and we study separately the cases of assigned displacement and assigned force acting on the handle of the experimental device.

\subsection{Hard device: Helmholtz ensemble}

To describe the system in thermal equilibrium in the case of assigned displacement, we consider the canonical partition function in the Helmholtz statistical ensemble $\mathscr{H}$. Due to the absence of non nearest neighborhood interactions, the chains energy depends only on the number $p$ of unfolded domains and not on the specific phase configuration $\chi$. As a result  (see SI and~\cite{fp:2019}) the partition function assumes the simple form
 \begin{equation}
 Z_{\mathscr{H}} = K_{\mathscr{H}}\sum_{p=0}^{n} \binom{n}{p} \, e^{-\frac{\beta k_m l \gamma n}{2}\left(\varepsilon_u \frac{p}{n}-(1+\alpha)\,\varepsilon_t\right)^2},
 \label{eq:19}
 \end{equation}
where $K_{\mathscr{H}}$ is a constant, taking into account also the kinetic energy, $\beta = 1/k_B T$, with $k_B$ the Boltzmann constant and $T$ the absolute temperature. The binomial coefficient gives the number of configurations of the chain with $p$ unfolded domains among the $n$ bistable  elements. \vspace{0.2 cm}

\noindent \textbf{Remark} \hspace{0.3 cm} We point out that (see SI) in order to obtain this analytical expression we assume that the two wells are extended beyond the spinodal point~\cite{efendiev:2010, fp:2019} and for fixed phase configuration $\chi_i$ we integrate each $\varepsilon_i$ in $\mathbb{R}$. In~\cite{fp:2019} the authors numerically showed that this approximation does not influence the energy minimization in the temperature regimes of interest for real experiments.\vspace{0.3 cm}

The Helmholtz free energy ($\mathcal{F}$) is, by definition,
\begin{equation}
\mathcal{F} = -\frac{1}{\beta} \ln Z_{\mathscr{H}}.
\label{henergy}
\end{equation}
Consequently (see SI) we can evaluate the expectation value of the force conjugated to the applied displacement $d$
\begin{equation}
\langle F \rangle = \frac{1}{L(1+\alpha)}\frac{\partial \mathcal{F}}{\partial \varepsilon_t}= k_m \gamma \,((1+\alpha)\,\varepsilon_t-\varepsilon_u\langle \bar{\chi} \rangle ),
\label{hforce}
\end{equation}
where 
\begin{equation}
\langle \bar{\chi} \rangle = \langle \bar{\chi} \rangle_{\mathscr{H}}(\beta, \varepsilon_t) = \frac{\displaystyle \sum_{p=0}^n\binom{n}{p}\frac{p}{n}\,e^{-\frac{\beta k_m l n \gamma}{2}\left(\varepsilon_u\frac{p}{n}-(1+\alpha)\,\varepsilon_t\right)^2}}{\displaystyle \sum_{p=0}^n\binom{n}{p}\,e^{-\frac{\beta k_m l n \gamma}{2}\left(\varepsilon_u\frac{p}{n}-(1+\alpha)\,\varepsilon_t\right)^2}}
\label{hchi}
\end{equation}
is the expectation value of the  fraction $\bar{\chi}$ of unfolded domains. After some manipulation (see again SI) we obtain the expectation value of the macromolecule strain 
\begin{equation}
\langle \varepsilon_m \rangle = \varepsilon_u\langle \bar{\chi} \rangle  + \gamma \left((1+\alpha)\varepsilon_t-\varepsilon_u\langle \bar{\chi} \rangle\right).
\label{hemet}
\end{equation}
Finally, by using \eqref{hforce} and \eqref{hemet} we obtain 
\begin{equation}
\langle F \rangle = k_m(\langle \varepsilon_m \rangle - \varepsilon_u \langle \bar{\chi} \rangle)
\label{hfem}
\end{equation}
by which we can study the effect of temperature and device stiffness (through the parameter $\gamma$) on the macromolecule. 

Observe that \eqref{hforce}, \eqref{hemet}, \eqref{hfem} are formally identical to the  equations \eqref{fepsilont}, \eqref{epsilontm}, and \eqref{Fo} obtained in the mechanical limit case,  with the only difference in the expression of the fraction, which is temperature dependent consistently with \eqref{hchi}.

\begin{figure}[!h]
\centering
  \includegraphics[width=0.45\textwidth]{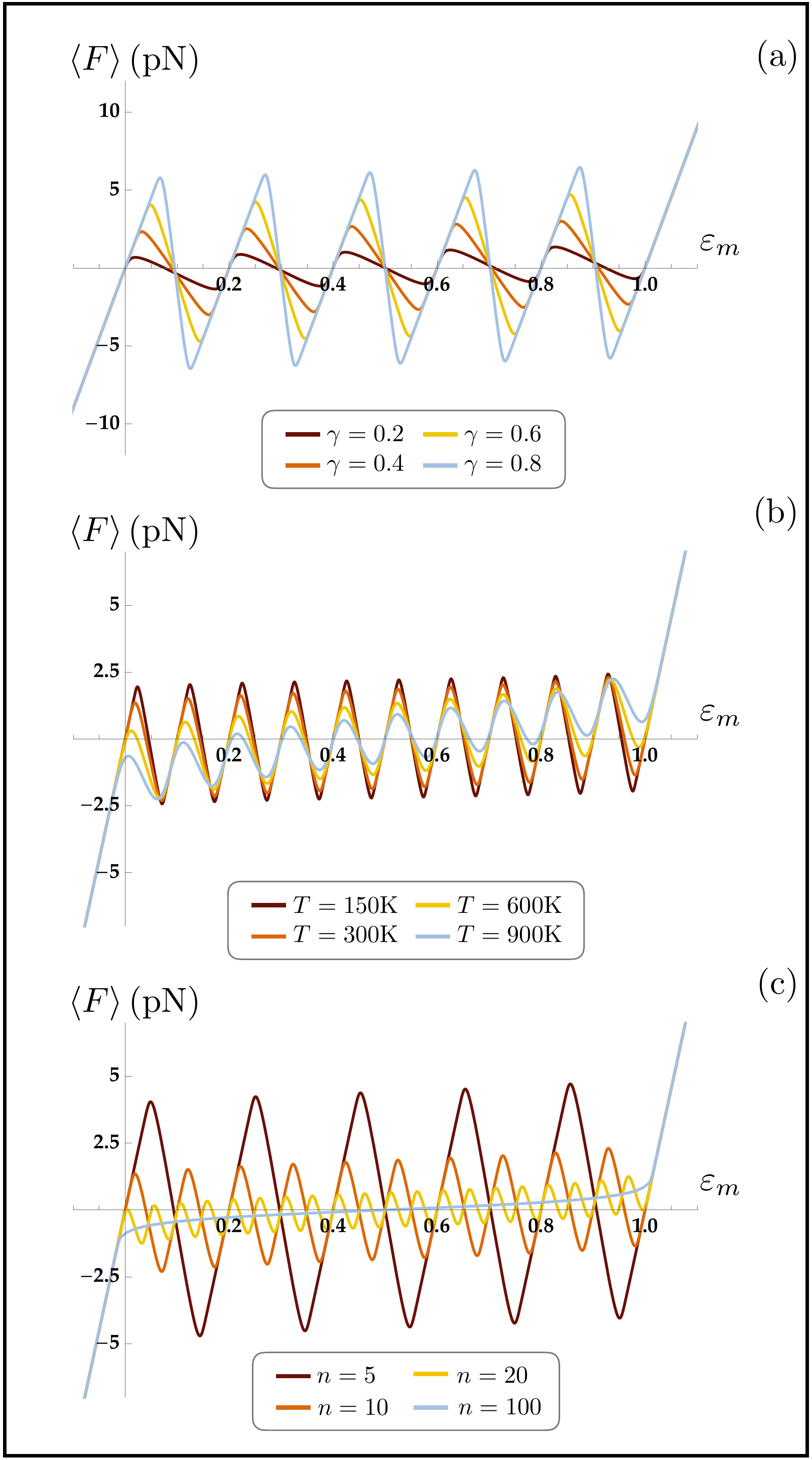}
  \caption{Hard device (Helmholtz ensemble) force-elongation diagrams. (a) Effect of $\gamma$, of temperature (b) and of the discreteness parameter $n$ on the strain of the macromolecule. Parameters:  $k_m=90\;\mbox{pN}$, $\alpha = 0.1$, $l=20\;\mbox{nm}$, $\varepsilon_u=1$, $n=5$ in (a) $n=10$ in (b).}
  \label{fig:helmholtz1}
\end{figure}

In Fig.~\ref{fig:helmholtz1} we show the influence of temperature,  device stiffness and number of elements $n$ of the chain on the unfolding behavior of the macromolecule.  A detailed interpretation of these results can be found in Section~\ref{sec:discussion}. 

\subsection{Soft device: Gibbs ensemble}

The partition functions of Gibbs ($\mathscr{G}$) and Helmholtz ($\mathscr{H}$) ensembles are related by a Laplace transform with force $F$ and displacement $d$ as conjugate variables~\cite{weiner:1983}. Thus, using \eqref{eq:5}, we have
\begin{equation}
Z_{\mathscr{G}}= \int Z_{\mathscr{H}}\,e^{\,\,\beta\, Fd}d d= (1+\alpha)L\int Z_{\mathscr{H}}\,e^{\beta \left(L(1+\alpha)\,\varepsilon_t\,F \right)}d\varepsilon_t.
\label{eq:25}
\end{equation}

A detailed calculation leads to a Gaussian integral whose solution is the partition function in the Gibbs canonical ensemble (see SI), that can be written as 
\begin{equation}
Z_{\mathscr{G}}= K_{\mathscr{G}}\sum_{p=0}^{n}\binom{n}{p}\,e^{\frac{\beta l n}{2 k_m \gamma}\left(F^2 + 2 k_m \gamma \varepsilon_u \frac{p}{n} F\right)}.
\label{pfg2}
\end{equation}

Again we used the simplifying result that due to the absence of non nearest neighborhood interactions the energy depends only on the number of unfolded elements.
In this case, it is possibile to evaluate explicitly this summation in order to obtain
\begin{equation}
Z_{\mathscr{G}}= K_{\mathscr{G}}\,\left(1+e^{\beta l \varepsilon_u F}\right)^ne^{\frac{\beta l n}{2 k_m \gamma}F^2}.
\label{pfg}
\end{equation}

The Gibbs free energy $\mathcal{G}$ is 
\begin{equation}
\mathcal{G}=-\frac{1}{\beta}\text{ln}\, Z_{\mathscr{G}}
\label{genergy}
\end{equation}
and, therefore, we can evaluate the expectation value of the total strain
\begin{equation}
\langle \varepsilon_t \rangle = \frac{1}{\beta L(1+\alpha)}\frac{1}{Z_{\mathscr{G}}}\frac{\partial Z_{\mathscr{G}}}{\partial F} = \frac{1}{(1+\alpha)}\left(\frac{F}{k_m\gamma}+\varepsilon_u\langle \bar{\chi} \rangle \right)
\label{gfet}
\end{equation}
where
\begin{equation}
\langle \bar{\chi} \rangle =\langle \bar{\chi} \rangle_{\mathscr{G}} (\beta, F)= \frac{e^{\,\beta l \varepsilon_u F}}{1+e^{\,\beta l \varepsilon_u F}}.
\label{gchi}
\end{equation}
\begin{figure}[!h]
	\centering
	 \includegraphics[width=0.45\textwidth]{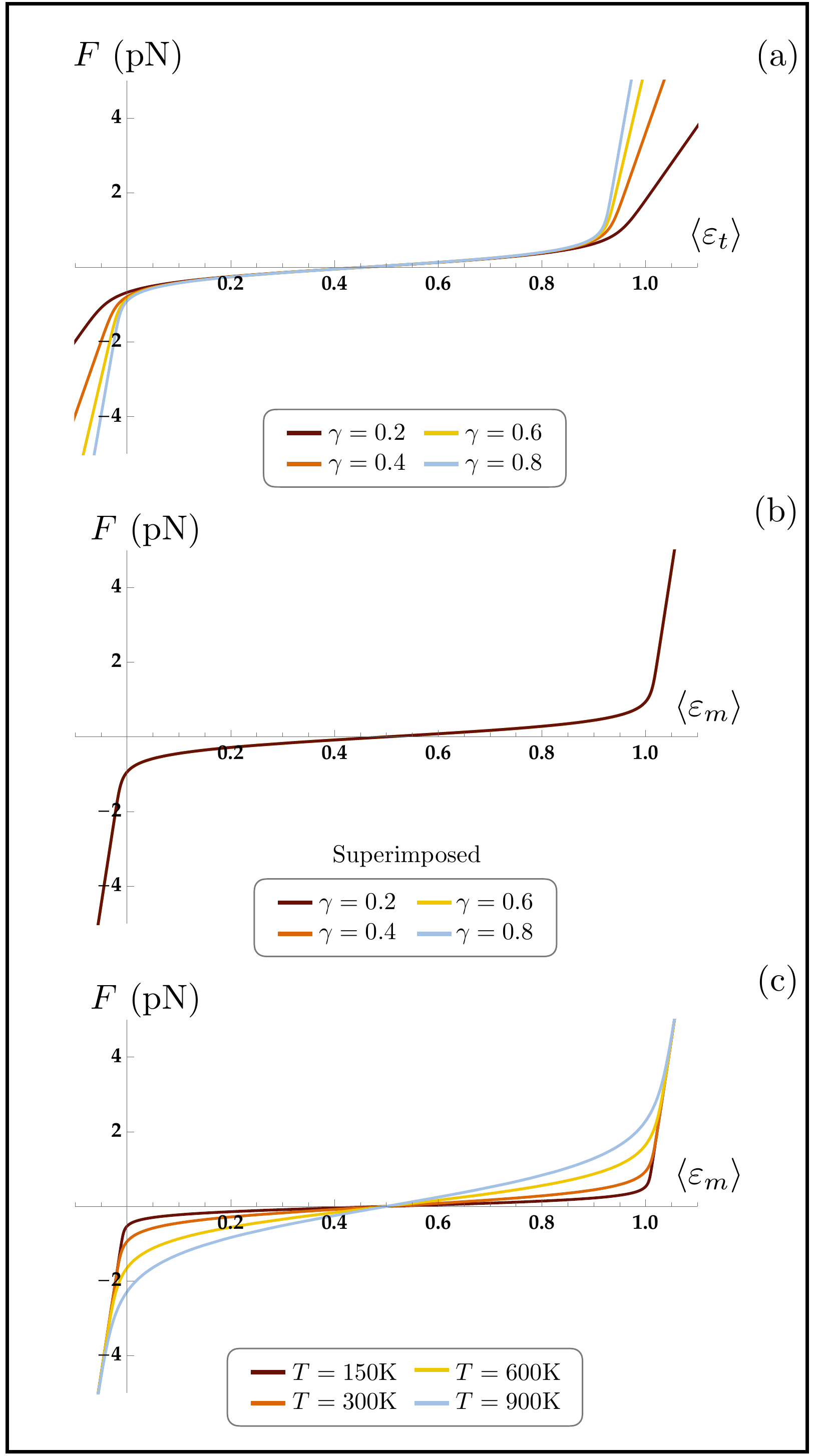}
	  \caption{Stress-strain curves of the Gibbs ensemble (assigned force). (a) Effect of $\gamma$. As $\gamma$ increases, the macromolecule becomes stiffer. (b) Effect of $\gamma$. As $\gamma$ increases, the force acting on the macromolecule is not affected. (c) Effect of the temperature $T$. As the temperature increases, the curves become steeper. The values used for the molecule properties are $n=5$, $k_m=90\;\mbox{pN}$, $\alpha = 0.1$, $l=20\;\mbox{nm}$, $\varepsilon_u=1$.}
	\label{fig:gibbs1}
\end{figure}
Also in the case of soft device (see SI~3) it is possible to show that  the total and the molecule strain are related by \eqref{hemet}, so that by using \eqref{gfet}  we obtain 
\begin{equation}
\langle \varepsilon_m \rangle = \frac{F}{k_m} +\varepsilon_u \langle \bar{\chi} \rangle
\label{gfem}.
\end{equation}
with a force strain molecular relation independent from the number  $n$ of elements of the chain. 

Once again, we point out the analogy among~\eqref{gfet},\eqref{gfem} and~\eqref{hemet} which are consistent with the analogous relations in the mechanical limit~\eqref{fepsilont},~\eqref{epsilontm} and ~\eqref{Fo} and in the Helmholtz ensemble ~\eqref{hforce},~\eqref{hfem} and ~\eqref{hemet} with the only difference given by \eqref{gchi}.

In Fig.~\ref{fig:gibbs1} we show the effects of temperature. It is important to remark that there is no influence of the device stiffness in the case of soft device regarding the force-strain macromolecule relation. We refer again to section \ref{sec:discussion} for a detailed discussion of these results.

\section{Thermodynamic limit}\label{sec:tl}

Many important biological molecules  undergoing conformational transitions, ({\it i.e.} titin~\cite{rief:1997} or DNA~\cite{busta:2000}) are constituted by a very large number of domains. Therefore, it is interesting to explore the thermodynamic limit behavior
 $n \to \infty$.  
 
Firstly, let us consider  the case of \textit{hard device}. In order to perform the thermodynamic limit we use the saddle point method that (see \cite{Z,fp:2019} and SI~5) delivers the expectation value of the unfolded fraction 
\begin{equation}
\langle \bar{\chi} \rangle\simeq \chi_c(\varepsilon_t),
\end{equation}
where $\chi_c$ is the solution of 
\begin{equation}
\text{ln}\left (\frac{x}{1-x}\right )+\varepsilon_u \beta k_m l \gamma \left(\varepsilon_u\,x-(1+\alpha)\,\varepsilon_t\right)= 0. 
\label{limitH}
\end{equation}

It is straightforward to see that, consistently with the results previously shown in the paper, we obtain the same form of the mechanical response of the macromolecules as in \eqref{hforce}, \eqref{hemet}, \eqref{hfem}.  The same equations can be extended also to the case of the \textit{soft device} thermodynamic limit, after observing that
 the phase fraction in \eqref{gchi} does not depend on $n$. 

We want now to extend previous results, regarding the equivalence of the molecule response under the hard and soft device in the thermodynamic limit to the case when non convex energies and the handle stiffness effect are considered. This result has been analytically shown in~\cite{winkler:2010, manca:2012} for flexible polymers in the case of convex energy. In the same papers the results have been numerically shown also in the case of a two-wells energy. Notice that, in the mechanical limit, the observed equivalence can be deduced following the approach in~\cite{puglisi:2000} and in~\cite{puglisi:2002} where the authors consider also  metastable configurations and hysteresis. On the other hand, this equivalence is not true when non local interactions are considered (see {\it e.g.}~\cite{lev:2004, puglisi:2007}).  

Since the force-elongation relation has the same expression in the two ensembles (see \eqref{hfem} and \eqref{gfem}), we prove the analytical equivalence in the thermodynamic limit by showing that the value of the unfolding fraction in \eqref{hchi} and \eqref{gchi} coincide. To this hand, since we used the Stirling approximation in the hard device case, we apply this formula also to \eqref{gchi} rewritten in the following form (derived from \eqref{pfg2} without evaluating the summation)
\begin{equation}
\langle \bar{\chi} \rangle = \frac{\displaystyle \sum_{p=0}^{n}\frac{p}{n}\,e^{\,\frac{\beta l n}{2 k_m \gamma}\left(F^2+2 k_m \varepsilon_u \gamma \frac{p}{n} F\right)}}{\displaystyle \sum_{p=0}^{n}\,e^{\,\frac{\beta l n}{2 k_m \gamma}\left(F^2+2 k_m \varepsilon_u \gamma \frac{p}{n} F\right)}}.
\end{equation}

Following an approach similar to the previous case (see SI Sect.~5), we found that the critical point $\chi_c$ is the solution of
\begin{equation}
\text{ln}\left (\frac{x}{1-x}\right )+ \beta  l \varepsilon_u F = 0. 
\end{equation}
By using \eqref{gfet} we  obtain again \eqref{limitH} proving the equivalence of the two ensembles in the thermodynamic limit. 

\section{Discussion}\label{sec:discussion}

We developed an exact {\it unified} mathematical model, in the framework of equilibrium Statistical Mechanics, quantifying the effect of the  handling device stiffness in stretching experiments on a chain of bi-stable elements. Among the many important examples one can think to SMFS tests on biomolecules. To fix the idea, in this paper we referred to AFM experiments on macromolecules constituted by a chain of domains (\textit{e.g.} $\alpha$-helix and $\beta$-sheets) undergoing conformational (folded $\rightarrow$ unfolded) transition. It is important to remark that the proposed framework can be extended to the case of mechanical molecular interactions~\cite{rosa:2004}.

In particular, following the approach in~\cite{fp:2019}, we considered the chain and the device microcantilever as a unique thermodynamical system. We analyzed both the assumptions of assigned displacement and applied force acting on the cantilever.  The former case is described by the so called Helmholtz ensemble, whereas the latter is described by the so called Gibbs ensemble, linked by an integral Laplace transform to the previous one. As we show, several limit cases of theoretical interest can be deduced by our general approach: the ideal hard and soft device (neglecting the device stiffness), the thermodynamic limit and the mechanical limit (neglecting entropy effects).

In particular, we deduced that in both statistical ensembles and in all considered limits, the molecular response can be formally described by the following relations:
\begin{equation}
\begin{array}{l}
 F  =  k_m \gamma \,((1+\alpha)\,\varepsilon_t-\varepsilon_u \bar{\chi} ),\vspace{0.2 cm}\\
 \varepsilon_m = \varepsilon_u \bar{\chi}   + \gamma \left((1+\alpha)\,\varepsilon_t-\varepsilon_u\bar{\chi} \right),
 \vspace{0.2 cm}\\
  F = k_m( \varepsilon_m  -  \varepsilon_u  \bar{\chi}),  
  \end{array}
  \label{vfa}
\end{equation}
where, with a slight abuse of notation, we identify the value of the variables with their expectation values depending on the specific case.

\begin{figure*}
	\centering
	 \includegraphics[width=0.94 \textwidth]{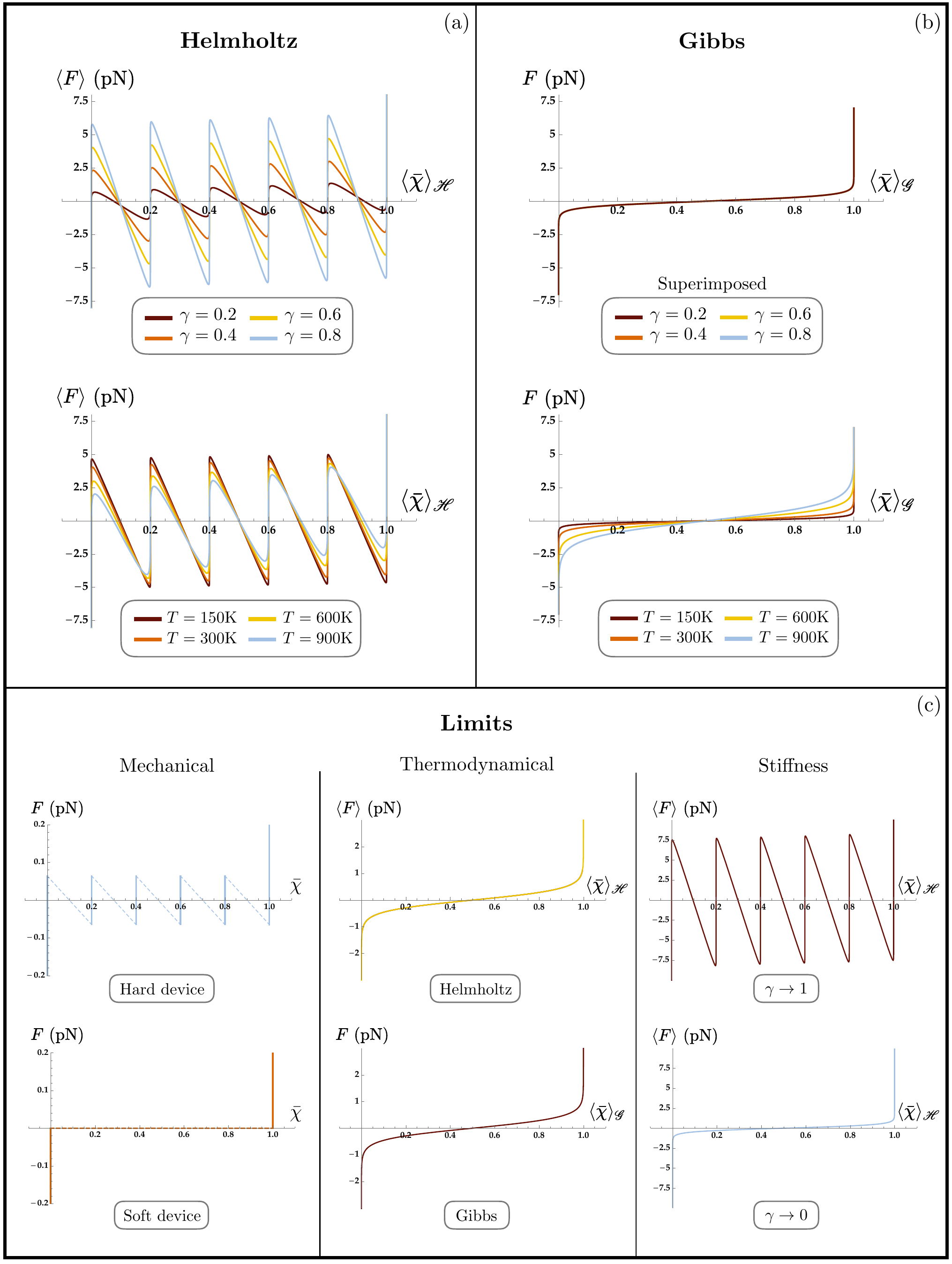}\vspace{0.1 cm}
	  \caption{Summary of the macromolecule behavior in terms of phase fraction evolution under different boundary conditions and limit regimes.}
	\label{fig:chis}
\end{figure*}

The  \textit{only} difference among all  considered possibilities lies on the expectation value of the unfolded fraction $\bar \chi$. Following this result, all the different behaviors analyzed in previous sections are described in Fig.~\ref{fig:chis}  with respect to the phase fraction evolutions during the molecule unfolding. 

The important influence on the mechanical response of the macromolecules when the total displacement is assigned, are shown in Fig.~\ref{fig:helmholtz1}a and Fig.~\ref{fig:chis}a.  As the device stiffness decreases, the unfolding force decreases and the behavior becomes more cooperative. It is important to remark that (see~\cite{fp:2019}) this can lead to huge overestimation or underestimation of the force thresholds of the macromolecule. 
In this perspective, we point out that an extended literature in the field neglects the stiffness effect and considers the ideal cases when the total macromolecule displacement is assigned ({\it ideal hard device}) or the force acting on the macromolecule is assigned ({\it ideal soft device}).

In passing, we observe that these limit behaviors can be deduced in our model by considering the limit of rigid device-molecule connection: $\gamma \to 1$. Indeed, as we show in SI, Sect.~6, in these ideal cases we obtain the same formal expressions in \eqref{vfa} with  phase fractions
\begin{equation}
\begin{array}{l}\displaystyle
 \langle \bar{\chi}^{ideal} \rangle_{\mathscr{H}} (\beta, \varepsilon_t)=\frac{\displaystyle\sum_{p=0}^{n}\binom{n}{p} \frac{p}{n}\,e^{-\frac{\beta k_m l n}{2}\left(\varepsilon_u\frac{p}{n}-\varepsilon_t\right)^2}}{\displaystyle\sum_{p=0}^{n}\binom{n}{p} e^{-\frac{\beta k_m l n}{2}\left(\varepsilon_u\frac{p}{n}-\varepsilon_t\right)^2}},
\vspace{0.4 cm}\\ \displaystyle
\langle \bar{\chi}^{ideal} \rangle_{\mathscr{G}} (\beta, F) = \frac{\displaystyle e^{\beta l  \varepsilon_u F}}{\displaystyle 1+e^{\beta l  \varepsilon_u F}}.
\end{array}
\end{equation}
The same expressions are obtained by the general model in the $\gamma\rightarrow1$ limit.

Moreover, it is interesting to observe that if we consider the hard device, when $\gamma \rightarrow 0$, the macromolecule response approaches the behavior of the ideal soft device (see Fig.~\ref{fig:chis}c down-right). Conversely, when $\gamma\rightarrow 1$ the macromolecular response coincide with the one of the ideal hard device (see Fig.~\ref{fig:chis}c up-right). Thus, by varying the stiffness ratio $\gamma$ all the ranges of behavior between the two limit cases can be attained.

The soft device boundary conditions are described in Fig.~\ref{fig:gibbs1} and Fig.~\ref{fig:chis}b. According with the experimental behavior  we may observe a monotonic transition force-elongation path. Interestingly, we obtain that in this case the macromolecule behavior is independent from $\gamma$  and $n$ (see Fig.~\ref{fig:gibbs1}b).

The temperature effect is shown in Fig.~\ref{fig:helmholtz1}b, Fig.~\ref{fig:gibbs1}c and in Fig.~\ref{fig:chis}a,b for the two ensembles. This should be compared with the limit regime when temperature effects are neglected, studied in Section~\ref{sec:zero}) and represented in Fig.~\ref{fig:energy} and Fig.~\ref{fig:chis}c left. Interestingly, while when we neglect temperature effects the unfolding transition corresponds to constant forces  thresholds, when temperature effects are considered we may observe a hardening behavior with the unfolding force growing with the unfolded percentage. This effect is in accordance with the experiments in~\cite{rief:1997} showing such a hardening behavior of a recombinant titin macromolecule with identical $\beta$-sheets domains. Observe that the hardening grows as the temperature grows.

The influence of the discreteness size is shown in Fig.~\ref{fig:helmholtz1}c. The force threshold decreases as the number of elements $n$ increases. The thermodynamic limit ($n\rightarrow\infty$) has been studied in section~\ref{sec:tl} and is represented in terms of phase fraction $\langle \bar \chi\rangle$ in Fig.~\ref{fig:chis}c center. Specifically, by extending the results in~\cite{manca:2012}, where the authors study generalizations of the freely jointed chain and of the worm-like chain models with extensible bonds, we demonstrated the equivalence  in the thermodynamic limit of the molecule response under hard and soft device. This result, was  numerically described  in~\cite{mancagiordano:2014}. 

In conclusion, we deduced a general framework, able to analytically describe in the rate-independent regime the importance of the stiffness  handle, considering also temperature effects. As we show, all other cases analyzed in the literature on SMFS experiments and models, such as thermodynamic limit~\cite{mancagiordano:2014,efendiev:2010}, mechanical limit~\cite{puglisi:2002} and ideal hard and soft devices~\cite{puglisi:2000,manca:2012,makarov:2009} can be obtained as limit cases of this general model.

%
\vspace{+0.5cm}
\paragraph{\textbf{Acknowledgements.}---}

GF and GP have been supported by the Italian Ministry MIUR-PRIN project `Mathematics of active materials: From mechanobiology to smart devices''. LB, GF and GP are supported by GNFM (INdAM). GF by INFN through the project ``QUANTUM'' and by the FFABR research grant.

%

\vspace{+0.5cm}
\paragraph{\textbf{Supporting Information.}---}
Supporting Information Available. 

%

\vspace{+0.5cm}
\paragraph{\textbf{Correspondence.}---} 
Correspondence and requests for materials should be addressed to G.F. (giuseppe.florio@poliba.it), G.P. (giuseppe.puglisi@poliba.it) or L.B. (luca.bellino@poliba.it)

%
\bibliography{LGGArxiv}

\clearpage
\newpage
\onecolumngrid
\section*{Supplementary Information for ``On the influence of device handle in single-molecule experiments''}\label{sec:suppmat}
 \renewcommand{\theequation}{SI\arabic{equation}}
\setcounter{page}{1}
\renewcommand{\theequation}
{SI-\arabic{equation}}

\setcounter{equation}{0}

\setcounter{figure}{0}

\noindent We report for the  reader convenience the analytical details of the model proposed  the paper. To this end we refer to the equations in the main paper using the same notation, whereas we denote with SI-(.) the equation (.) of the Supporting Information document. Moreover according with the main paper notation we denote with the subscripts $m$, $d$, and $t$ to denote variables referred to the molecule, the device and the total (molecule plus device) system, respectively.

\section{Hamiltonian of the system and basic definitions}

\noindent The system is composed of $n$ mass points with mass $m$ connected by bistable springs with modulus $k_m$ and a loading device with mass $M$ represented as an $n+1$ spring with modulus $k_d$. The Hamiltonian function can be written as the sum of kinetic and elastic energy 
\begin{equation}
H = E_{K} + V_{t} = \sum_{i=1}^{n}\frac{1}{2 m}p_i^2+\frac{1}{2M}p^2_{n+1}+\sum_{i=1}^{n}\frac{1}{2}k_m l (\varepsilon_i-\varepsilon_u\chi_i)^2+\frac{1}{2}k_d \alpha L \varepsilon_d^2 
\label{H}
\end{equation}
where $\varepsilon_u$ is the reference strain of the unfolded configuration, 
 $l = L/n$ is the reference length of each element, $\alpha L \varepsilon_d$ is the elongation of the device and $p_i$ are the momentum of the $i$-th oscillator. Here $\chi_i$ is an internal variable that can assume values $0$ or $1$ if the $I$-the element is \textit{folded} or \textit{unfolded}, respectively. We consider different boundary conditions acting on the device: assigned displacement $d$ (\textit{hard device}) or assigned force $F$ (\textit{soft device}). In the framework of equilibrium Statistical Mechanics, these two cases are described by the Helmholtz and Gibbs ensembles, respectively. The ideal hard and soft device, with assigned displacements and force, respectively, acting directly on the molecule are obtained as limit systems. Finally we consider the mechanical limit when entropic effects can be neglected and the  thermodynamic limit when the number $n$ of elements diverges.
 
 The total displacement can be expressed as
\begin{equation}
d = \sum_{i=1}^{n}l\varepsilon_i+\alpha L \varepsilon_d = L(\varepsilon_m +\alpha \varepsilon_d)
\label{eq:2}
\end{equation}
where $\varepsilon_m$ is the average strain of the macromolecule 
\begin{equation}
\varepsilon_m=\frac{1}{n}\sum_{i=1}^{n}\varepsilon_i.
\label{average}
\end{equation}
The relation between $\varepsilon_m$and the total strain $\varepsilon_t$
\begin{equation}
\varepsilon_t=\frac{d}{L(1+\alpha)}\Rightarrow (1+\alpha)\, \varepsilon_t=\varepsilon_m+\alpha\, \varepsilon_d.
\label{eq:3}
\end{equation}
By using \eqref{eq:3} we can express the device strain as
\[
\alpha\, \varepsilon_d=(1+\alpha)\, \varepsilon_t-\varepsilon_m.
\]
The equilibrium condition requires a constant force $F$:
$$k_m(\varepsilon_i-\varepsilon_u\chi_i)=k_d\,\varepsilon_d.$$
By averaging with respect to $I$ we then get the relation between the macromolecule and the device strain
\[
k_d \,\varepsilon_d = k_m(\varepsilon_m-\varepsilon_u \bar{\chi})
\]
where $\bar{\chi}=\frac{\sum_{i=1}^n \chi_i}{n}$ is the fraction of unfolded domains.  %
After introducing the non-dimensional parameter  
\[
\gamma=\frac{k_d}{k_d+\alpha\, k_p}  \in [0,1],
\]
measuring the relative device vs total stiffness, by using (\ref{eq:3}) we get
\begin{equation}
\varepsilon_m=(1+\alpha)\gamma\, \varepsilon_t+(1-\gamma)\,\varepsilon_u\bar{\chi}.
\label{mecstress}
\end{equation}
%

\section{Helmholtz Ensemble}
\noindent Consider first the case of hard device, when the total displacement $d$ is applied to the instrument. In this case we have to consider the partition function in the Helmholtz ensemble  defined as 
\[
Z_{\mathscr{H}}=(l^n \alpha L )\sum_{\boldsymbol{\chi}}\int_{\mathbb{R}^{2(n+1)}} e^{-\beta H}\delta \biggl(\sum_{i=1}^{n}l\varepsilon_i+\alpha L \varepsilon_d - d\biggr) \prod_{i}^{n}dp_i \,dp_{d} \prod_{i}^{n} l\, d\varepsilon_{i} \,(\alpha L)\,d\varepsilon_d 
\]
where $\beta = 1/k_BT$, $k_B$ being the Boltzmann constant, $T$ the absolute temperature and $\boldsymbol{\chi}= \{\chi_{1},\dots ,\chi_{n}\}\in \{0,1\}^{n}$ is the vector denoting the phase (folded of unfolded) configuration. For the sake of simplicity  we drop the domain of the vector spin variable $\boldsymbol{\chi}$. We  used the delta function
 to enforce the displacement constraint (\ref{eq:2}).

%
%
We can separate the contributions to $Z_{\mathscr{H}}$ of the kinetic energy and of the potential energy, so that we can split up the integral over the momenta and over the strains, respectively: 
\[
Z_{\mathscr{H}}= \alpha L l^n \int_{\mathbb{R}^{(n+1)}} e^{-\beta E_k}\prod_{i=1}^{n} dp_i\,dp_{d}  \sum_{\boldsymbol{\chi}}\int_{\mathbb{R}^{(n+1)}} e^{-\beta V_t}\delta \biggl(\sum_{i=1}^{n}l\varepsilon_i+\alpha L \varepsilon_d - d\biggr)\prod_{i=1}^{n} d\varepsilon_i\,d\varepsilon_d.
\]
We solve the Gaussian integral over the momenta to obtain 
\[
Z_{\mathscr{H}}=A \alpha L l^n  \sum_{\boldsymbol{\chi}} \int_{\mathbb{R}^{(n+1)}}e^{-\beta \bigl(\sum_{i}\frac{1}{2}k_m l (\varepsilon_i-\varepsilon_u\chi_i)^2+\frac{1}{2}\frac{k_d}{\alpha L}(\alpha L \varepsilon_d)^2\bigr)}\delta \biggl(\sum_{i}l\varepsilon_i+\alpha L \varepsilon_d - d\biggr)\prod_{i} d\varepsilon_i\,d\varepsilon_d,
\]
where
\begin{equation}
A=(2\pi)^{(n+1)/2}\biggl(\frac{m}{\beta }\biggr)^{n/2} \biggl(\frac{M}{\beta}\biggr)^{1/2}.
\label{A}
\end{equation}

We can also integrate out the free variable $\varepsilon_d$ to obtain 
\[
Z_{\mathscr{H}}=A\,l^n \sum_{\boldsymbol{\chi}} \int_{\mathbb{R}^{n}}e^{-\beta \Bigl(\sum_{i}\frac{1}{2}k_m l (\varepsilon_i-\varepsilon_u\chi_i)^2+\frac{1}{2}\frac{k_d}{\alpha L}\bigl(\sum_{i}l\varepsilon_i-d\bigr)^2\Bigr)}\prod_{i} d\varepsilon_i.
\]
By using (\ref{eq:3}) we get
\begin{equation}
\label{ZZ}
Z_{\mathscr{H}}=A\,l^n\sum_{\boldsymbol{\chi}} \int_{\mathbb{R}^{n}}e^{-\frac{\beta l k_m}{2} \Bigl(\sum_{i} (\varepsilon_i-\varepsilon_u\chi_i)^2+\eta n\,\bigl(\frac{1}{n}\sum_{i}\varepsilon_i-(1+\alpha)\varepsilon_t\bigr)^2\Bigr)}\prod_{i} d\varepsilon_i,
\end{equation}
where 
\begin{equation}
\eta = \frac{k_d}{\alpha k_m}=\frac{\gamma}{1-\gamma} \quad\text{with}\quad \eta \in [0,+\infty[.
\label{eq:16}
\end{equation}
In order to solve the Gaussian integrals, we rearrange the exponent in the partition function as follows:
\begin{eqnarray}
&&-\frac{\beta l k_m}{2} \left(\left(1+\frac{\eta}{n}\right)\sum_{i=1}^{n} \varepsilon_i^2+\frac{\eta}{n}\sum_{i,j=1, i\ne j}^{n} \varepsilon_i \varepsilon_j-2\sum_{i=1}^{n}(\varepsilon_u \chi_i + \eta (1+\alpha)\varepsilon_t)\varepsilon_i+\eta n (1+\alpha)^2\varepsilon_t^2+\varepsilon_u^2\sum_{i=1}^{n}\chi_i^2\right)=\nonumber\\
&&-\frac{1}{2}\boldsymbol{A}\boldsymbol{\varepsilon} \cdot \boldsymbol{\varepsilon}+ \boldsymbol{b}\cdot\boldsymbol{\varepsilon}+C
\end{eqnarray}
where we have introduced 
\begin{eqnarray}
&&\boldsymbol{A} = \beta k_m l\begin{pmatrix}
1+\frac{\eta}{n} & \frac{\eta}{n} & \dots & \frac{\eta}{n} \\
\frac{\eta}{n}  & \ddots & &\vdots \\
\vdots & & \ddots& \vdots\\
\frac{\eta}{n} & \dots &\dots & 1+\frac{\eta}{n}\\
\end{pmatrix}, \vspace{0.2 cm}\\ 
&&\boldsymbol{b}= \{\beta k_m l \bigl(\varepsilon_u \chi_1 + \eta \,(1+\alpha)\varepsilon_t\bigr),\dots, \beta k_m l \bigl(\varepsilon_u \chi_n + \eta \,(1+\alpha)\varepsilon_t\bigr) \}^T, \vspace{0.2 cm} \\ 
&&\boldsymbol{\varepsilon} = \{\varepsilon_1,\dots,\varepsilon_n\}^T,
\end{eqnarray}
%
%
and
\[
C=-\frac{\beta k_m l}{2}\left(n\,\eta (1+\alpha)^2\varepsilon_t^2+\varepsilon_u^2\sum_{i=1}^{n}\chi_i^2\right)
\]
is a constant energy term. The  Gaussian integration of quadratic functions can be solved explicitly (see {\it e.g.} \cite{Z} giving
\[
\int_{\mathbb{R}^{n}} e^{-\frac{1}{2}\boldsymbol{A}\boldsymbol{\varepsilon} \cdot \boldsymbol{\varepsilon}+ \boldsymbol{b}\cdot\boldsymbol{\varepsilon}+C}d\boldsymbol{\varepsilon} = \sqrt{\frac{(2\pi)^n}{\mbox{det}\boldsymbol{A}}}\,e^{\frac{1}{2}\boldsymbol{A}^{-1}\boldsymbol{b}\cdot\boldsymbol{b}+C}.
\]

Thus, we obtain 
\[
Z_{\mathscr{H}}=K_{\mathscr{H}} \sum_{\boldsymbol{\chi}}  \,e^{\frac{\beta l k_m}{2}\Bigl(\sum_{i}\bigl(\varepsilon_u \chi_i + \eta (1+\alpha)\varepsilon_t\bigr)-\frac{\gamma}{n}\Bigl(\sum_{i}\bigl(\varepsilon_u \chi_i + \eta (1+\alpha)\varepsilon_t\bigr)\Bigr)^2-\varepsilon_u^2\sum_i\chi_i^2-\eta n (1+\alpha)^2\varepsilon_t^2\Bigr)}
\label{eq:19}
\]
with
\[
K_{\mathscr{H}} = A\,l^n\biggl(\frac{2\pi}{\beta k_m l}\biggr)^{n/2}(1-\gamma)^{1/2}.
\]
We observe that, due to the absence of non-local energy terms, all solutions with the same unfolded fraction $
\bar{\chi}$ are characterized by the same energy. As a result the partition function describing the chain and the apparatus as a whole is 
\begin{equation}
Z_{\mathscr{H}}=K_{\mathscr{H}}\sum_{p=0}^n\binom{n}{p}\,e^{-\frac{\beta k_m l n \gamma}{2}\bigl(\frac{p}{n}\varepsilon_u-(1+\alpha)\varepsilon_t\bigr)^2}.
\label{partitionhelm}
\end{equation}
Notice that the binomial coefficient gives the number of iso-energetic configurations for fixed value of $p$.

We then deduce that the Helmholtz free energy is given by
\[
\mathcal{F}=-\frac{1}{\beta}\ln Z_{\mathscr{H}}
\]
and the expectation value of the force can be obtained as 
\begin{equation}
\label{FF}
\langle F \rangle = \frac{1}{L(1+\alpha)}\frac{\partial \mathcal{F}}{\partial \varepsilon_t}=-\frac{1}{\beta L (1+\alpha)}\frac{1}{Z_{\mathscr{H}}}\frac{\partial Z_{\mathscr{H}}}{\partial \varepsilon_t}.
\end{equation}

Observe that the force-strain relation can be written in the same form of Eq.({\color{red}11}) of the main paper
\begin{equation}
\langle F \rangle = k_m \gamma (\varepsilon_t(1+\alpha)-\varepsilon_u \langle \bar{\chi} \rangle)
\label{etFhelm}
\end{equation}
after introducing the (temperature  dependent) expectation value of the unfolded fraction
\begin{equation}
\langle \bar{\chi} \rangle = \langle \bar{\chi} \rangle_{\mathscr{H}}(\beta, \varepsilon_t) = \frac{\displaystyle\sum_{p=0}^n\binom{n}{p}\frac{p}{n}\,e^{-\frac{\beta k_m l n \gamma}{2}\bigl(\frac{p}{n}\varepsilon_u-(1+\alpha)\varepsilon_t\bigr)^2}}{\displaystyle\sum_{p=0}^n\binom{n}{p}\,e^{-\frac{\beta k_m l n \gamma}{2}\bigl(\frac{p}{n}\varepsilon_u-(1+\alpha)\varepsilon_t\bigr)^2}}.
\label{chiH}
\end{equation}

In order to evaluate the expectation value of the macromolecule strain, it is convenient to start from the expression \eqref{ZZ}. We have
\[
\langle \varepsilon_m \rangle = \frac{A}{Z_{\mathscr{H}}} \sum_{\boldsymbol{\chi}}\int_{\mathbb{R}^{n}}\biggl(\frac{1}{n}\sum_{i=1}^n\varepsilon_i\biggr)\,e^{-\frac{\beta l k_m}{2} \Bigl(\sum_{i} (\varepsilon_i-\varepsilon_u\chi_i)^2+\eta n\,(\frac{1}{n}\sum_{i}\varepsilon_i-(1+\alpha)\varepsilon_t\bigr)^2\Bigr)}\prod_{i=1}^n l\, d\varepsilon_i,
\]
where $A$ is given by (\ref{A}).
It is straightforward to show that
\begin{equation}
\frac{1}{L(1+\alpha)}\frac{1}{Z_{\mathscr{H}}}\frac{\partial Z_{\mathscr{H}}}{\partial \varepsilon_t} = -\beta\, k_m \frac{\gamma}{1-\gamma}((1+\alpha)\varepsilon_t-\langle \varepsilon_m \rangle),
\label{media}
\end{equation}
and, thus,
\begin{equation}
\langle \varepsilon_m \rangle=(1+\alpha) \varepsilon_t-\frac{1-\gamma}{k_m \gamma }\left(-\frac{1}{\beta L(1+\alpha)}\frac{1}{Z_{\mathscr{H}}}\frac{\partial Z_{\mathscr{H}}}{\partial \varepsilon_t} \right)=\varepsilon_u \langle \bar{\chi} \rangle + \gamma \bigl((1+\alpha)\varepsilon_t-\varepsilon_u\langle \bar{\chi} \rangle\bigr).
\label{EPSM}
\end{equation}
%
with the same form of the mechanical limit in Eq.({\color{red}12}).
Finally, we have
\begin{equation}
\langle F \rangle = k_m \left(\langle \varepsilon_m\rangle- \varepsilon_u\langle \bar{\chi} \rangle\right),
\end{equation}
again respecting the results in Eq.({\color{red}14}) of the mechanical limit, with the variation due to the expectation value of $\bar \chi$ in \eqref{chiH}.

%
\section{Gibbs Ensemble}
\noindent Consider now the case of assigned force (soft device). 
The partition function for the Gibbs canonical ensemble is
\[
Z_{\mathscr{G}}=\alpha L \sum_{\boldsymbol{\chi}}\int_{\mathbb{R}^{2(n+1)}}e^{-\beta \Bigl(H-F\bigl(\sum_{i=1}^{n}l\varepsilon_i+\alpha L \varepsilon_d\bigr)\Bigr)}\prod_{i}^n dp_{i}\,dp_d \prod_{i}^n \, l\, d\varepsilon_i \,d\varepsilon_d.
\]
where the Hamiltonian is defined in \eqref{H}. We obtain
\[
\begin{array}{lll}
Z_{\mathscr{G}}&=&\displaystyle A (\alpha L\,l^n )\sum_{\boldsymbol{\chi}}\int_{\mathbb{R}^{(n+1)}}e^{-\beta \Bigl(\frac{1}{2}\sum_{i=1}^{n}\bigl(k_m l (\varepsilon_i - \varepsilon_u \chi_i)^2 - F l\varepsilon_i \bigr) + \frac{1}{2}\frac{k_d}{\alpha L}(\alpha L \varepsilon_d)^2  -F\alpha L \varepsilon_d\Bigr)}\prod_{i=1}^{n}d\varepsilon_i\,d\varepsilon_d \\
&=&\displaystyle A(\alpha L\,l^n )\sum_{\boldsymbol{\chi}}\int_{\mathbb{R}^{(n+1)}}e^{-\frac{\beta l k_m}{2} \Bigl(\sum_{i}\bigl((\varepsilon_i-\varepsilon_u \chi_i)^2-\frac{2F}{k_m}\varepsilon_i\bigr)+\frac{1-\gamma}{\gamma} n(\alpha \varepsilon_d)^2-\frac{2F}{k_m}n\alpha \varepsilon_d\Bigr)}\prod_{i}d\varepsilon_i\,d\varepsilon_d \nonumber\\
&=& A(\alpha L\,l^n ) \, I_m \,I_d
\end{array}
\]
where $A$ has the same value \eqref{A} obtained in the case of assigned displacement, we used Eq. (\ref{eq:16}), $I_m$ and $I_d$ correspond to the integration with respect to the $\varepsilon_i$ and $\varepsilon_d$, respectively. We easily obtain
\begin{equation}
I_d=C_{\mathscr{G}}\,e^{\frac{\beta l n}{2 k_m \gamma}(1-\gamma)\,F^2}
\label{eq:49}
\end{equation}
where we have defined the constant
\[
C_{\mathscr{G}}=\frac{1}{\alpha}\biggl(\frac{2\pi (1-\gamma)}{\beta k_m l \gamma n}\biggr)^{1/2}.
\]
On the other hand, the integral $I_m$ can be rewritten as
\[
I_m=\sum_{\boldsymbol{\chi}}\prod_{i=1}^n\int_{\mathbb{R}}e^{-\frac{\beta l k_m}{2} \left((\varepsilon_i-\varepsilon_u \chi_i)^2-\frac{2F}{k_m}\varepsilon_i \right)}d\varepsilon_i =  \biggl(\frac{2\pi}{\beta k_m l}\biggr)^{n/2}\,\sum_{\boldsymbol{\chi}}\prod_{i=1}^n e^{\frac{\beta l }{2 k_m}\,\bigl(F^2+F\,2 k_m \varepsilon_u \chi_i\bigr)}.
\]
Also in this case, we may observe that, due to the absence of non local energy terms, the energy of the solutions with the same unfolded fraction is invariant with respect to the permutation of the elements. Thus, we obtain the analytic expression
\begin{eqnarray}
I_m  &=&\biggl(\frac{2\pi}{\beta k_m l}\biggr)^{n/2}\sum_{p=0}^n \binom{n}{p}\Bigl(e^{\frac{\beta l}{2 k_m}\,F^2}\Bigr)^{n-p}\Bigl(e^{\frac{\beta l}{2 k_m}(F^2+F\,2 k_m \varepsilon_u)}\Bigr)^{p}\nonumber\\
&=&\biggl(\frac{2\pi}{\beta k_m l}\biggr)^{n/2}\sum_{p=0}^n \binom{n}{p}\,e^{\frac{\beta l n}{2 k_m}\,\bigl(F^2+F\,2 k_m \varepsilon_u \frac{p}{n}\bigr)}=\biggl(\frac{2\pi}{\beta k_m l}\biggr)^{n/2}\,e^{\frac{\beta l n}{2k_m}F^2 }\left(1+e^{\beta l \varepsilon_u F}\right)^n.
\label{eq:55}
\end{eqnarray}
%
Finally, we find the partition function in the Gibbs ensemble:
\begin{equation}
Z_{\mathscr{G}}=K_{\mathscr{G}}\,e^{\frac{\beta l n}{2 k_m \gamma}\,F^2}\left(1+e^{\beta l \varepsilon_u F}\right)^n,
\label{eq:partfgibbs}
\end{equation}
where 
\[
K_{\mathscr{G}}=A\,(\alpha L\,l^n )\,\biggl(\frac{2\pi}{\beta k_m l}\biggr)^{n/2}\,C_{\mathscr{G}}.
\]

Based on this result we can deduce the constitutive force-strain relation in the case of assigned force.  By using the definition of  average strain \eqref{average}, we get
\begin{equation}
\langle \varepsilon_m \rangle = \frac{A\,(\alpha L\,l^n )\,I_d}{Z_{\mathscr{G}}}\,\sum_{\boldsymbol{\chi}}\int_{\mathbb{R}^{n}}\left(\frac{1}{n}\sum_{i=1}^{n}\varepsilon_i \right)\,e^{-\frac{\beta l k_m}{2} \sum_{i}\left((\varepsilon_i-\varepsilon_u \chi_i)^2-\frac{2F}{k_m}\varepsilon_i\right)}\prod_{i=1}^{n}d\varepsilon_i.
\label{BBB}
\end{equation}
This can be rewritten as
\[
\langle \varepsilon_m \rangle = \frac{A\,(\alpha L\,l^n )\,I_d}{Z_{\mathscr{G}}}\,\sum_{\boldsymbol{\chi}}\frac{1}{n}\sum_{i=1}^{n}\left(\prod_{j=1}^{i-1}\int_{\mathbb{R}^{i-1}}e^{-\tilde{\beta}h(\varepsilon_j,\chi_j)}d\varepsilon_j\,\int_{\mathbb{R}}  \varepsilon_i\,e^{-\tilde{\beta}h(\varepsilon_i,\chi_i)}d\varepsilon_i\prod_{k=i+1}^{n}\int_{\mathbb{R}^{n-i-1}}e^{-\tilde{\beta}h(\varepsilon_k,\chi_k)}d\varepsilon_k\right)
\]
where $\tilde{\beta}=\beta l k_m/2$ and 
\[
h(\varepsilon,\chi)=\left((\varepsilon-\varepsilon_u \chi)^2-\frac{2F}{k_m}\varepsilon\right).
\]
Thus, we have a product of simple Gaussian integrals (the integral over $\varepsilon_i$ requires an integration by parts). The solution can be written as
\begin{eqnarray}
\langle \varepsilon_m \rangle &=& \left(\frac{2\pi}{k_m \beta l}\right)^{n/2}\frac{A\,(\alpha L\,l^n )\,I_d}{Z_{\mathscr{G}}}\sum_{\boldsymbol{\chi}}\frac{1}{n}\sum_{i=1}^{n}\,
\Biggl(\frac{1}{k_m}\left(F+k_m \varepsilon_u \chi_i \right) e^{\frac{\beta l}{2 k_m}(F^2+2 F k_m \varepsilon_u \chi_i)}\times\nonumber\\
&\times&\prod_{j=1}^{i-1}e^{\frac{\beta l}{2 k_m}(F^2+2 F k_m  \varepsilon_u \chi_j)}\times \prod_{k=i+1}^{n}e^{\frac{\beta l}{2 k_m}(F^2+2 F k_m  \varepsilon_u \chi_k)}\Biggr)\nonumber\\
&=&\left(\frac{2\pi}{k_m \beta l}\right)^{n/2} \frac{A\,(\alpha L\,l^n )\,I_d}{Z_{\mathscr{G}}}\sum_{\boldsymbol{\chi}}\frac{1}{n}\sum_{i=1}^{n}\left(\frac{1}{k_m}\left(F+k_m \varepsilon_u \chi_i \right)\prod_{j=1}^{n}e^{\frac{\beta l}{2 k_m}(F^2+2 F k_m \varepsilon_u \chi_j)} \right).
\label{integralegigante}
\end{eqnarray}
By simplifying \eqref{integralegigante} we get
\begin{eqnarray}
\langle \varepsilon_m \rangle &=&\left(\frac{2\pi}{k_m \beta l}\right)^{n/2} \frac{A\,(\alpha L\,l^n )\,I_d}{Z_{\mathscr{G}}}\sum_{\boldsymbol{\chi}}\frac{1}{n}\sum_{i=1}^{n}\left(\frac{1}{k_m}\left(F+k_m \varepsilon_u \chi_i \right)\prod_{j=1}^{n}e^{\frac{\beta l}{2 k_m}(F^2+2 F k_m \varepsilon_u \chi_j)} \right)\nonumber\\
&=&\left(\frac{2\pi}{k_m \beta l}\right)^{n/2}\frac{A\,(\alpha L\,l^n )\,I_d}{Z_{\mathscr{G}}}\sum_{\boldsymbol{\chi}}\left(\left(\frac{F}{k_m}+ \varepsilon_u \frac{1}{n}\sum_{i=1}^{n}\chi_i \right)\prod_{j=1}^{n}e^{\frac{\beta l}{2 k_m}(F^2+2 F k_m  \varepsilon_u \chi_j)} \right)\nonumber\\
&=&\left(\frac{2\pi}{k_m \beta l}\right)^{n/2}\frac{A\,(\alpha L\,l^n )\,I_d}{Z_{\mathscr{G}}}\left(\sum_{p=0}^n\binom{n}{p}\left(\frac{F}{k_m}+\varepsilon_u\frac{p}{n}\right)\,e^{\frac{\beta l n}{2 k_m}\,\bigl(F^2+F\,2 k_m \varepsilon_u \frac{p}{n}\bigr)}\right).
\end{eqnarray}
where in the last equality we   followed the same procedure used in \eqref{eq:55}.
Finally, using the final form of the partition function \eqref{eq:partfgibbs} and the integrals $I_m$, $I_d$ we obtain 
\begin{equation}
\langle \varepsilon_m \rangle = \frac{F}{k_m}+\varepsilon_u \langle \bar{\chi} \rangle
\label{kmFgib}
\end{equation}
that again has the same form of the molecular response Eq.({\color{red}14}) in the purely mechanical approximation, but  in this case we consider the expectation value of the unfolded fraction $\langle \bar{\chi} \rangle$ in the Gibbs ensemble 
\begin{equation}
\langle \bar{\chi} \rangle = \langle \bar{\chi} \rangle_{\mathscr{G}}(\beta, F)=\frac{\displaystyle\sum_{p=0}^n\binom{n}{p}\frac{p}{n}e^{\frac{\beta l n}{2 k_m}\,\bigl(F^2+F\,2 k_m \varepsilon_u \frac{p}{n}\bigr)}}{\displaystyle\sum_{p=0}^n\binom{n}{p}e^{\frac{\beta l n}{2 k_m}\,\bigl(F^2+F\,2 k_m \varepsilon_u \frac{p}{n}\bigr)}}=\frac{e^{ l \beta \varepsilon_u F}}{1+e^{ l \beta \varepsilon_u F}}.
\label{chigibbs}
\end{equation}

By definition, the Gibbs free energy is
\[
\mathcal{G}=-\frac{1}{\beta}\ln Z_{\mathscr{G}}
\]
and the expectation value of the total strain of the system, which is the variable conjugated to the force, can be obtained as 
\begin{equation}
\langle \varepsilon_t \rangle = \frac{1}{\beta L(1+\alpha)}\frac{1}{Z_{\mathscr{G}}}\frac{\partial}{\partial F}Z_{\mathscr{G}}.
\label{defet}
\end{equation}
This leads to
\begin{equation}
\langle \varepsilon_t \rangle = \frac{1}{(1+\alpha)}\biggl(\frac{F}{k_m\gamma}+\varepsilon_u \langle \bar{\chi} \rangle\biggr),
\label{etF}
\end{equation}
where we have used \eqref{chigibbs}, that has the same form of that again has the same form of  Eq.({\color{red}11}).

 From \eqref{kmFgib} and \eqref{etF} we can obtain the relation between $\langle \varepsilon_t \rangle$ and $\langle \varepsilon_m \rangle$
\begin{equation}
\langle \varepsilon_m \rangle=\varepsilon_u \langle \bar{\chi} \rangle + \gamma \bigl((1+\alpha)\langle\varepsilon_t\rangle-\varepsilon_u\langle \bar{\chi} \rangle\bigr).
\end{equation}
 consistent with Eq.({\color{red}11}) of the mechanical limit.
\section{From Helmholtz to Gibbs ensembles: Laplace Transform}

\noindent As well known \cite{weiner:1983}, the partition functions in the Gibbs and Helmholtz ensembles are connected by a Laplace transform with the force $F$ and the total displacement $d$ as conjugate variables. From \eqref{partitionhelm} we have
\begin{eqnarray}
\int_\mathbb{R} Z_{\mathscr{H}}\,e^{\beta \, Fd} dd&=& (1+\alpha)L\int_\mathbb{R} Z_{\mathscr{H}}\,e^{\beta \bigl(F\, L(1+\alpha)\varepsilon_t \bigr)}d\varepsilon_t\nonumber\\
&=&K_{\mathscr{H}} (1+\alpha)L\sum_{p=0}^{n}\binom{n}{p}\int_\mathbb{R} e^{-\beta l n\Bigl(\frac{k_m \gamma}{2}\bigl(\frac{p}{n}\varepsilon_u-(1+\alpha)\varepsilon_t\bigr)^2-F (1+\alpha)\varepsilon_t\Bigr)}d\varepsilon_t\nonumber\\
&=&K_{\mathscr{H}}L\left(\frac{2\pi}{k_m l n \beta\gamma}\right)^{1/2}\sum_{p=0}^{n}\binom{n}{p}\,e^{\frac{\beta l n}{2 k_m \gamma}\Bigl(F^2 + F\, 2 k_m \gamma \varepsilon_u \frac{p}{n}\Bigr)}\nonumber\\
&=&K_{\mathscr{G}}\,e^{\frac{\beta l n}{2 k_m \gamma}\,F^2}\left(1+e^{\beta l \varepsilon_u F}\right)^n=Z_{\mathscr{G}},
\label{partgibs}
\end{eqnarray}
which is exactly the result obtained in \eqref{eq:partfgibbs}. The other quantities $\langle \varepsilon_t \rangle$ and $\langle \varepsilon_m \rangle$ in the Gibbs ensemble can be obtained accordingly.

\section{Thermodynamical limit}

\noindent In this section we show how to evaluate the expression of the phase fraction expression \eqref{chiH} in the thermodynamical limit by using the saddle point method \cite{Z}. According to previous discussion the dependence of the response from temperature, device stiffness and discreteness parameter $n$ is measured by  the expectation value of the unfolded fraction, being the other expectation values of mechanical variable related by the same equations Eq.({\color{red}11}), Eq.({\color{red}12}), Eq.({\color{red}14}).

Since in the Gibbs ensemble, the formula \eqref{chigibbs} does not depend on $n$ the thermodynamical limit behavior coincides with the one of systems with finite discreteness. We then need to study only the limit  the Helmholtz ensemble $\langle \bar{\chi} \rangle$ in \eqref{chiH}.  To this end, we start considering the function $f$ defined as
\begin{equation}
f(\varepsilon_t) = \sum_{p=0}^n\binom{n}{p}\,e^{-\frac{\beta k_m l n \gamma}{2}\left(\frac{p}{n}\varepsilon_u-\left(1+\alpha\right)\varepsilon_t\right)^2}.
\label{tlimit1}
\end{equation}
Using the Stirling approximation, $n!\sim \left (\frac{n}{e}\right )^n \sqrt{2\pi n}$ for $n\gg1$, \eqref{tlimit1} can be written as 
\[
f(\varepsilon_t) \simeq \frac{1}{\sqrt{2 \pi n}}\sum_{p=0}^n \, \sqrt{\frac{1}{p/n(1-p/n)}}\,\displaystyle  e^{n\, \text{ln}\,n-p\, \text{ln}\, p-(n-p)\, \text{ln}(n-p)-\frac{\beta k_m l n \gamma}{2}\left(\frac{p}{n}\varepsilon_u-\left(1+\alpha\right)\varepsilon_t\right)^2},
\]
where we  considered both $n$ and $p$ large.
To deduce the themodynamic limit, let us introduce the variable  $x=p/n$. In the limit of large $n$ we obtain
\[
f(\varepsilon_t) \simeq \sqrt{\frac{n}{2 \pi}}\int_{0}^1\, \sqrt{\frac{1}{x(1-x)}}\,\displaystyle  e^{-n \left( S(x) +\frac{\beta k_m l  \gamma}{2}\left(x\,\varepsilon_u-\left(1+\alpha\right)\varepsilon_t\right)^2\right )}\text{d}x
\]
where we  defined the (entropy) function
\[
S(x) = x\, \text{ln}x + (1-x)\text{ln}\, (1-x).
\]
Finally, for large $n$ we can apply the saddle point approximation. We search for the minimum of the function
\[
 S(x)+\frac{\beta k_m l \gamma}{2}\left(\,\varepsilon_u  x-  (1+\alpha)\varepsilon_t\right)^2
 \]
which can be found solving the equation
\begin{equation}
\text{ln}\left (\frac{x}{1-x}\right )+\varepsilon_u \beta k_m l \gamma \left(\,\varepsilon_u x-(1+\alpha) \varepsilon_t\right)= 0. 
\label{hlimitt}
\end{equation}
It is easy to see that there exists only one solution $\chi_{c}$ in the interval $]0,1[$. Thus, we can solve the integral with the saddle point method by considering the expansion around $\chi_c$ up to the second order as it follows: 
\[
f(\varepsilon_t) \simeq \sqrt{\frac{n}{2 \pi}}\int_{0}^1\, \sqrt{\frac{1}{\chi_{c}(1-\chi_{c})}}\,  e^{-n \left(S(\chi_{c}) -\frac{\beta k_m l n \gamma}{2}\left(\,\varepsilon_u\chi_{c}-\left(1+\alpha\right)\varepsilon_t\right)^2-\frac{1}{2}\left(S''(\chi_{c})+\beta l k_m \gamma \varepsilon_u^2\right)\left(x-\chi_{c}\right)^2\right)}\text{d}x.
\]
By substituting the variable $y=\sqrt{n}(x-\chi_{c})$  we get
\[
f(\varepsilon_t) \simeq \sqrt{\frac{1}{2\pi \chi_{c}(1-\chi_{c})}}\,e^{-n \left(S(\chi_{c}) -\frac{\beta k_m l n \gamma}{2}\left(\,\varepsilon_u \chi_{c}-\left(1+\alpha\right)\varepsilon_t\right)^2\right)} \int_{-\sqrt{n}\,\chi_{c}}^{\sqrt{n}\,\chi_{c}} e^{-\frac{1}{2}\left(S''(\chi_{c})+\beta l k_m \gamma \varepsilon_u^2\right)y^2}\text{d}y.
\]
In the limit $n\rightarrow\infty$, we obtain
\[
f(\varepsilon_t) \sim \frac{\displaystyle e^{-n\left(S(\chi_{c})+\frac{\beta k_m l \gamma}{2}\left( \varepsilon_u\chi_{c}-\varepsilon_t(1+\alpha)\right)^2\right)}}{\displaystyle \sqrt{1+\beta k_m l \gamma \varepsilon_u \chi_{c}(1-\chi_{c})}}.
\]
Similarly, we can show that
\[
g(\varepsilon_t)=\sum_{p=0}^n\binom{n}{p}\,\frac{p}{n}\,e^{-\frac{\beta k_m l n \gamma}{2}\left(\frac{p}{n}\varepsilon_u-\left(1+\alpha\right)\varepsilon_t\right)^2} \sim \frac{\displaystyle \chi_{c}\, e^{-n\left(S(\chi_{c})+\frac{\beta k_m l \gamma}{2}\left(\chi_{c} \varepsilon_u-(1+\alpha)\varepsilon_t\right)^2\right)}}{\displaystyle \sqrt{1+\beta k_m l \gamma \varepsilon_u \chi_{c}(1-\chi_{c})}}.
\]
Finally, we get
\begin{equation}
\langle \bar{\chi} \rangle  =\frac{g(\varepsilon_t)}{f(\varepsilon_t)} \sim \chi_{c}(\varepsilon_t).
\end{equation}

We can now easily extend previous results in \cite{winkler:2010} and \cite{manca:2012} on the equivalence of the results obtained from Helmholtz and Gibbs ensembles in the thermodynamical limit also for systems with non convex energies of interest in this paper. We can rewrite the expectation value of the total strain in the Gibbs ensemble, see \eqref{etF}, as
\begin{equation}
\langle \varepsilon_t \rangle = \rho(F).
\label{eq:etF2}
\end{equation}
On the other hand, we want to show that the expectation value $\langle F \rangle$ in the Helmholtz ensemble converges to $F$ in the thermodynamical limit. Since in both cases we found the same strain-force relations with the only difference in the expectation value of the phase fraction, we only need to show that the two expressions 
\eqref{chiH} and \eqref{chigibbs} attain the same limit as $n$ diverges. This can be done by  simply verifying that $\langle\bar{\chi}\rangle_{\mathscr{G}}$ is the only solution of \eqref{hlimitt}: 
\begin{equation}
\text{ln}\left (\frac{\langle\bar{\chi}\rangle_{\mathscr{G}}}{1-\langle\bar{\chi}\rangle_{\mathscr{G}}}\right )+\varepsilon_u \beta k_m l \gamma \left(\varepsilon_u\,\langle\bar{\chi}\rangle_{\mathscr{G}}-(1+\alpha)\langle \varepsilon_t \rangle \right)=F l \beta \varepsilon_u-F l \beta \varepsilon_u= 0. 
\label{hlimitt2}
\end{equation}

%

%
\section{Ideal Cases}
\noindent In this section we consider the \textit{ideal} cases 
Tipically considered in the literature, when the device effect is neglected and the displacement ({\it ideal hard device})
Or the force ({\it ideal soft device}) are directly applied to the unfolding molecule. In this case
$
\varepsilon_m \equiv  \varepsilon_t$ and the Hamiltonian  is  
\begin{equation}
H^{ideal} =\sum_{i=1}^{n}\frac{1}{2 m}p_i^2+ \frac{1}{2}k_m l\sum_{i=1}^{n}(\varepsilon_i-\varepsilon_u \chi_i)^2 .
\label{hideal}
\end{equation}
%

\subsection{Ideal Helmholtz ensemble}

\noindent Using \eqref{hideal}, the partition function in the Helmholtz ensemble for the ideal case is 
\begin{equation}
Z_{\mathscr{H}}^{ideal} = \sum_{\boldsymbol{\chi}}l^n\int_{\mathbb{R}^{2n}}e^{-\beta H^{ideal}}\delta\left(l\sum_{i=1}^n\varepsilon_i-d\right)\prod_{i=1}^{n}dp_i\prod_{i=1}^{n}d\varepsilon_i.
\end{equation}
The integrals over the momenta result in the constant 
\[
A_{\mathscr{H}}^{ideal}=l^n (2\pi)^{\frac{n}{2}}\left(\frac{m}{\beta}\right)^{\frac{n}{2}}.
\]
The constraint on the total displacement is imposed by the Dirac delta as it follows: 
\begin{eqnarray}
Z_{\mathscr{H}}^{ideal} &=& A_{\mathscr{H}}^{ideal} \sum_{\boldsymbol{\chi}} \int_{\mathbb{R}^{n}} e^{\,-\beta\left(\frac{k_m l}{2}\sum_{i=1}^{n-1}\left(\varepsilon_i-\varepsilon_u \chi_i\right)^2+\frac{k_m l}{2}\left(\varepsilon_n-\varepsilon_u \chi_n\right)^2\right)}\delta\left(l\sum_{i=1}^{n-1}\varepsilon_i+l\varepsilon_n-d\right)\prod_{i=1}^{n}d\varepsilon_i\nonumber\\
&=&A_{\mathscr{H}}^{ideal} \sum_{\boldsymbol{\chi}} \int_{\mathbb{R}^{n-1}} e^{\,-\frac{\beta k_m l}{2}\left(\sum_{i=1}^{n-1}\left(\varepsilon_i-\varepsilon_u \chi_i\right)^2+\left(\sum_{i=1}^{n-1}\varepsilon_i-\varepsilon_u\chi_n-n\,\varepsilon_m\right)^2\right)}\prod_{i=1}^{n-1}d\varepsilon_i\nonumber\\
&=&A_{\mathscr{H}}^{ideal} \sum_{\boldsymbol{\chi}} \int_{\mathbb{R}^{n-1}} e^{-\frac{1}{2}\boldsymbol{A}\boldsymbol{\varepsilon} \cdot \boldsymbol{\varepsilon}+ \boldsymbol{b}\cdot\boldsymbol{\varepsilon}+C}\prod_{i=1}^{n-1}d\varepsilon_i
\end{eqnarray}
%
where we have introduced 
\begin{eqnarray}
&&\boldsymbol{A} = \beta k_m l\begin{pmatrix}
2 & 1 & \dots & 1 \\
1 & 2 & &\vdots \\
\vdots & & \ddots& \vdots\\
1 & \dots &\dots & 2\\
\end{pmatrix}, \\ 
&&\boldsymbol{b}= \{\beta k_m l \left(\varepsilon_u \chi_1 +\varepsilon_u \chi_n + n\,\varepsilon_m\right),...,\beta k_m l \left(\varepsilon_u \chi_{n-1} +\varepsilon_u \chi_n + n\,\varepsilon_m\right) \}^T, \\ 
&&\boldsymbol{\varepsilon} = \{\varepsilon_1,...,\varepsilon_{n-1}\}^T,
\end{eqnarray}
%
and
\[
C=\varepsilon_u^2\sum_{i=1}^{n-1}\chi_i^2+\varepsilon_u^2\chi_n^2+n^2\varepsilon_m^2+2\varepsilon_u\chi_n\varepsilon_m\,n.
\]
The Gaussian integration can be solved as before in the general case with the presence of the device \cite{Z}. We obtain 
\[
Z_{\mathscr{H}}^{ideal} = K_{\mathscr{H}}^{ideal} \sum_{\boldsymbol{\chi}} \,e^{\frac{\beta k_m l }{2}\left(\sum_{i=1}^{n-1}\left(\varepsilon_u\chi_i+\varepsilon_u\chi_n+ n\,\varepsilon_m\right)^2 - \frac{1}{n}\left(\sum_{i=1}^{n-1}\left(\varepsilon_u\chi_i+\varepsilon_u\chi_n+ n\,\varepsilon_m\right)\right)^2-\varepsilon_u^2\sum_{i=1}^{n-1}\chi_i^2-\varepsilon_u^2\chi_n^2-n^2\varepsilon_m^2-2\varepsilon_u\chi_n \varepsilon_m n \right)}
\]
with
\[
K_{\mathscr{H}}^{ideal} = A_{\mathscr{H}}^{ideal} \sqrt{\frac{(2\pi)^{n-1}}{(\beta k_m l )^{n-1}\,n}}.
\]
Finally, we obtain the  partition function for the ideal case in the Helmholtz ensemble: 
\[
Z_{\mathscr{H}}^{ideal} = K_{\mathscr{H}}\sum_{p=0}^n\binom{n}{p}\,e^{-\frac{\beta k_m l n}{2}\left(\frac{p}{n}\varepsilon_u-\varepsilon_t\right)^2}.
\]
Using a procedure analogous to the general case, we deduce the formula for the expectation value of the unfolded fraction in the ideal case:
\begin{equation}
\langle \bar{\chi} ^{ideal}\rangle = \langle \bar{\chi}^{ideal} \rangle_{\mathscr{H}}(\beta, \varepsilon_t) = \frac{\displaystyle\sum_{p=0}^n\binom{n}{p}\frac{p}{n}\,e^{-\frac{\beta k_m l n }{2}\left(\frac{p}{n}\varepsilon_u-\varepsilon_t\right)^2}}{\displaystyle\sum_{p=0}^n\binom{n}{p}\,e^{-\frac{\beta k_m l n }{2}\left(\frac{p}{n}\varepsilon_u-\varepsilon_t\right)^2}}.
\label{chiHideal}
\end{equation}
%

\subsection{Ideal Gibbs Ensemble}

\noindent If we apply a fixed force at the end of the chain of $n$ bistable elements without considering the measuring device we obtain the case of \textit{ideal soft device}. By using \eqref{hideal} we can write the partition function in the Gibbs ensemble as
\[
Z_{\mathscr{G}}^{ideal} = \sum_{\boldsymbol{\chi}}\int_{\mathbb{R}^{2n}}e^{-\beta \left( H^{ideal}- F l \sum_{i=1}^{n} \varepsilon_i \right)} \prod_{i=1}^{n}dp_i \prod_{i=1}^{n}l\, d\varepsilon_i.
\]
As in the Helmholtz ensemble the integral over the momenta give the constant 
\[
A_{\mathscr{G}}^{ideal}=A_{\mathscr{H}}^{ideal}=l^n (2\pi)^{\frac{n}{2}}\left(\frac{m}{\beta}\right)^{\frac{n}{2}}.
\]
The integrals over the strains can be rewritten as 
\begin{eqnarray}
Z_{\mathscr{G}}^{ideal} &=& A_{\mathscr{G}}^{ideal} \sum_{\boldsymbol{\chi}}\int_{\mathbb{R}^{n}}e^{-\frac{\beta k_m l }{2}\sum_{i=1}^{n}\left(\left(\varepsilon_i-\varepsilon_u\chi_i\right)^2-\frac{2F}{k_m}\varepsilon_i\right)}\prod_{i=1}^{n}d\varepsilon_i\nonumber\\
&=&A_{\mathscr{G}}^{ideal} \sum_{\boldsymbol{\chi}}\prod_{i=1}^{n}\int_{\mathbb{R}}e^{-\frac{\beta k_m l }{2}\left(\left(\varepsilon_i-\varepsilon_u\chi_i\right)^2-\frac{2F}{k_m}\varepsilon_i\right)}d\varepsilon_i.
\end{eqnarray}
%


The solution can be obtained exactly as in Sect.~3. We have 
\begin{equation}
Z_{\mathscr{G}}^{ideal} = K_{\mathscr{G}}^{ideal}\sum_{p=0}^n \binom{n}{p}\,e^{\,\frac{\beta l n}{2 k_m}\,\bigl(F^2+2 k_m \varepsilon_u \frac{p}{n}F\bigr)} = K_{\mathscr{G}}^{ideal} \,e^{\frac{\beta l n}{2 k_m \gamma}F^2}\Bigl(1+e^{l \beta \varepsilon_u F}\Bigr)^n.
\label{zidealg}
\end{equation}

From \eqref{zidealg} we can obtain, as in the previous cases, the expectation value of the strain of the molecule and the expectation value of the unfolded fraction in the ideal case 
\[
\langle \varepsilon_m \rangle = \frac{F}{k_m}+\varepsilon_u \langle \bar{\chi}^{ideal} \rangle,
\]
\[
\langle \bar{\chi}^{ideal} \rangle = \langle \bar{\chi}^{ideal} \rangle_{\mathscr{G}}(\beta, F) = \frac{e^{\, \beta l \varepsilon_u F}}{1+e^{\, \beta l \varepsilon_u F}}.
\]

\end{document}